\documentstyle[11pt]{article}
\normalbaselines
\begin{document}
\bibliographystyle{unsrt}

\def\bea*{\begin{eqnarray*}}
\def\eea*{\end{eqnarray*}}
\def\ba{\begin{array}}
\def\ea{\end{array}}
\count1=1
\def\be{\ifnum \count1=0 $$ \else \begin{equation}\fi}
\def\ee{\ifnum\count1=0 $$ \else \end{equation}\fi}
\def\ele(#1){\ifnum\count1=0 \eqno({\bf #1}) $$ \else \label{#1}\end{equation}\fi}
\def\req(#1){\ifnum\count1=0 {\bf #1}\else \ref{#1}\fi}
\def\bea(#1){\ifnum \count1=0   $$ \begin{array}{#1}
\else \begin{equation} \begin{array}{#1} \fi}
\def\eea{\ifnum \count1=0 \end{array} $$
\else  \end{array}\end{equation}\fi}
\def\elea(#1){\ifnum \count1=0 \end{array}\label{#1}\eqno({\bf #1}) $$
\else\end{array}\label{#1}\end{equation}\fi}
\def\cit(#1){
\ifnum\count1=0 {\bf #1} \cite{#1} \else 
\cite{#1}\fi}
\def\bibit(#1){\ifnum\count1=0 \bibitem{#1} [#1    ] \else \bibitem{#1}\fi}
\def\ds{\displaystyle}
\def\hb{\hfill\break}
\def\comment#1{\hb {***** {\em #1} *****}\hb }

\newcommand{\TZ}{\hbox{T\hspace{-5pt}T}}
\newcommand{\MZ}{\hbox{I\hspace{-2pt}M}}
\newcommand{\ZZ}{\hbox{Z\hspace{-3pt}Z}}
\newcommand{\NZ}{\hbox{I\hspace{-2pt}N}}
\newcommand{\RZ}{\hbox{I\hspace{-2pt}R}}
\newcommand{\CZ}{\,\hbox{I\hspace{-6pt}C}}
\newcommand{\PZ}{\hbox{I\hspace{-2pt}P}}
\newcommand{\QZ}{\hbox{I\hspace{-6pt}Q}}
\newcommand{\HZ}{\hbox{I\hspace{-2pt}H}}
\newcommand{\EZ}{\hbox{I\hspace{-2pt}E}}
\newcommand{\GZ}{\,\hbox{l\hspace{-5pt}G}}
\newcommand{\DZ}{\, \hbox{l\hspace{-5pt}D}}

\vbox{\vspace{38mm}}
\begin{center}
{\LARGE \bf Heisenberg and Modular Invariance  of
\\[2mm] N=2 Conformal Field Theory 
 }\\[5mm]

Shi-shyr Roan
\footnote{
Supported in part by the NSC grant of
Taiwan.}  \\{\it
Institute of Mathematics
\\ Academia Sinica \\  Taipei , Taiwan \\ (e-mail:
maroan@ccvax.sinica.edu.tw)} \\[5mm]
\end{center}

\begin{abstract} 
We present a theta function representation of the
twisted characters for the rational N=2
superconformal field theory, and discuss
the Jacobi-form like functional properties of 
these characters for a fixed central charge 
under the action of a finite Heisenberg group and
modular transformations.

\end{abstract}

\vfill
\eject

\section{Introduction}

Superconformal algebras have been studied for a long time in physical 
literature. 
The importance is above all due to the fact that the superconformal 
algebra describes the underling symmetries of Superstring theory. 
A superconformal algebra is a simple Lie 
superalgebra (over $\CZ$) 
spanned by modes of 
a finite family of local fields, which contains
the Virasoro  and some other odd , even ones,
such that the coefficients  of the operator
product expansions are linear combinations of
fields  in the family and their derivatives. 
This infinite-dimensional  algebra possesses a
very rich mathematical structure with  vast
applications on different branches in
physics and mathematics.  Nevertheless, only
until recent years, some rigorously systematic
study of conformal superalgebras from the
mathematical point of view has been  worked out
by groups of mathematicians and physicists, and
profound mathematical  structures have been found
to further enhance our understanding of  the 
original physical content (see, for instance
 \cite{FSST, Ka, Wa} and references therein).
However, despite the many theories of 
conformal algebras, the N=2 conformal theory 
has a peculiar nature which defines
a class of its own, other than due to its
applicable physical contents of the theory 
(see \cite{DG, DVV, EOTY} references therein),
also the mathematical reason on the intimate 
relation of special features of the N=2
superalgebra 
with K\"{a}hler geometry which has been a
main core in the analytical study of projective
algebraic manifolds.   Hence it would be expected
that  one of the 
important aspects of  N=2 conformal
theory will be on the geometrical understanding
 of complex manifolds, especially the
application on their topological invariants. A
notable example would be the  elliptic genus of
$c_1=0$ manifolds, (see for instance
\cite{ KYY, W} and references
therein), which amounts to the  so called
Gepner-model construction of the rational
conformal theory
\cite{Gep}. As the mathematical structure of
elliptic genus has now been well understood by
works of the Hirzebruch school \cite{HBJ, H}, it is
expected that a rigorous treatment of this
topological invariant in terms of the N=2 conformal
theory would be appeared as, or in a further
complete form than, the one given in
\cite{R}. However,  due to the length of 
presentation and also the nature of the contents,
in this paper 
we shall  only discuss the
representation-theoretical aspect of the N=2
conformal  algebra and leave its application on
topology of  manifolds as a separate problem, 
which will be treated in another work
\cite{R00}. In this note, we shall
study both the
quantitative and qualitative 
properties of character functions  of the
rational N=2 conformal theory. The unitary N$=2$
superconformal characters have been analyzed by 
several groups of physicists
\cite{BFK, D, Ki, M, RY}.  While working on
characters, they are usually regarded as the
formal Laurent series; however a proper
theta-form expression of the character is
expected for  the discussion of symmetries among
the characters. By incorporating the 
characteristic of theta functions into the twisted
currents of the algebra,  we obtain  a theta
function (with characteristic) presentation of
the characters of  unitary irreducible highest
weight modules (HWM) of N=2 conformal algebra for
different sectors,   (see Theorem 1 of the
content). These  explicit forms of characters in
terms of the theta function, which seem to be
unknown before to the best of the author's
knowledge, do provide a convenient formulation 
for discussing the global  functional
properties through the classical elliptic function
theory. Deceptively simple as it may appear, this
is an crucial observation since it implies that in
this form, the symmetries between
characters of  the rational N=2 conformal theory
for a fixed central charge are clearly revealed
and the later applications are expected to
achieve in the study of the geometry of
manifolds.

The paper is organized as follows. 
In Sect. 2, we give a brief introduction of the
elliptic, modular functional properties of the
Heisenberg and modular transformations. We also
summarize a few known facts about the
Jacobi-form like nature of theta functions (with
characteristic) in the elliptic function theory.
In Sect. 3, we first summarize some basic facts
about the  N=2 conformal algebra and the discrete
series of unitary irreducible HWM with the
central charge $c<3$. Then we derive the
theta form representation  
of  characters of these HWM for all different
sectors, which enables us to further pursue
their  qualitative functional properties, instead
of using only the Laurent series
expression. In Sect. 4, we study the symmetries
among these characters 
of HWM in the discrete series with a fixed
central element. 
With the elliptic and modular properties, the 
characters
constitute an irreducible  representation of  
a central extension of some Jacobi-group, which
composes of a finite Heisenberg group 
and a certain modular group.  We consider this
part as one of the most delicate aspects of this
paper. For this reason we describe, as precisely
as possible, the procedure of constructing the
irreducible decomposition of 
Heisenberg group with a certain explicit basis, by
which a matrix realization of modular
transformations is given and the
modular action on the
irreducible Heisenberg-submodules is
clearly shown. The detailed description will be
stated in Theorem 2 of the paper. Finally, as it
is known ( cf.,  for instance \cite{GR} and
references therein, ) that  superconformal
algebras have a certain  geometrical superspace 
realization in which  generators are presented as
super-vector fields (i.e., derivations) of 
super-spacetime. A such kind of
 geometrization for the usual
Virasoro algebra has been a well-known process,
by which  the  symmetry content of  the algebra
is better revealed. Hence in the
last section as an appendix, we  in particular
recall and describe a detailed account of a
geometrical realization of the N=2 conformal
algebra as 1-dimensional vector fields  with 2
supersymmetric extensions. This geometrical
setting might offer a means of realizing the 
superconformal symmetry, which has been the 
focus of the present work.

{\bf Convention .} In this paper, 
$\RZ, \CZ$ will denote 
the field of real, complex numbers respectively, 
and 
$\HZ$ the upper  half plane $  \{ \tau \in
\CZ \ |\ {\rm Im} \tau > 0\}$; $\RZ^*=
\RZ -\{0\}$, $\CZ^* =\CZ -\{0\}$. By a
representation of a group
$G$ on a (complex) vector space 
$V$, we shall 
always mean a linear right-action of $G$ on $V$, 
$$
R: V \times G \longrightarrow V \ , \ \ 
(v , g) \mapsto v \cdot g : = R(v, g) \ .
$$
A $R$-eigenvector with the eigen-character $\rho$ 
is an element $v \in V$ satisfying the relation: 
$$
v \cdot g = \rho(g) v \ , \ \ \forall g \in G \ .
$$

\section{Elliptic and Modular
Functional Properties }
Denote ${\cal O}(\CZ \times \HZ)$ the 
vector space of holomorphic functions of 
$\CZ \times \HZ$. 
In this section, we shall give a brief review on some basic elliptic 
and modular functional 
properties of $\CZ \times \HZ$, 
which will be needed for our later discussions.  
Here the ellipticity 
and modularity we refer to are the following 
standard actions of $\RZ^2$, $SL_2(\RZ)$ on 
$\CZ \times \HZ$ respectively,
$$
\begin{array}{cl}
\RZ^2 \times (\CZ \times \HZ)  \longrightarrow \CZ \times \HZ \ , 
& ( v, (z, \tau) ) \mapsto ( z+ v_1 \tau + v_2 ,
\tau) ,  \  v= ( v_1, v_2) \ , 
\\
 SL_2(\RZ) \times (\CZ \times \HZ) \longrightarrow \CZ \times \HZ , 
&(M, (z, \tau) ) \mapsto ( 
\frac{z}{C\tau+D}, \frac{A\tau +B}{C\tau+D} )  \ , \ 
M= \left( \begin{array}{cc}
A&B\\ 
C&D
\end{array}
\right) .
\end{array}
$$ 
For the convenience of notations, in this section 
the letters, $v_1, v_2$, will always  
denote 
the coordinates of a vector $v$ in $\RZ^2$;
$A, B, C, D$ are the entries of a matrix 
$M $ of $SL_2(\RZ)$. Denote 
$\RZ^2*SL_2(\RZ)$ the semi-product of $\RZ^2$
and $SL_2(\RZ)$  where the conjugation of $M \in
SL_2(\RZ)$ on $v
\in
\RZ^2$ is given   by  $M^{-1} \cdot v \cdot M =
vM$ (the matrix product). The above expressions 
give rise to a geometrical action of $\RZ^2 *
SL_2(\RZ)$ on
$\CZ\times \HZ$. One would like to lift this 
geometrical 
action to one on ${\cal O}(\CZ \times
\HZ)$. The elliptic part is as follows.  Let
$(\RZ^2, <,>)$ be the skew-symmetric form, $ <v
,   v'>: = v_1v_2'- v_2v_1'$ for $ v , v' \in
\RZ^2$, and denote
$\CZ^*_1 = \{ \alpha \in \CZ \ | \ | \alpha | = 1
\}$.
For $d \in \RZ^*$, $\GZ
(d)= \CZ_1^* \times \RZ^2$ is the Heisenberg group
(of  index $d$) with the group law, 
$$
(\alpha, v )\cdot (\alpha', v' )  = (\alpha \alpha'
e^{- 2  d \pi \sqrt{-1}  v_2v_1'}, v+v' ) \ , \ \ 
{\rm for } \ \alpha, \alpha' \in \CZ^*_1 \ , \ \ 
v , \ v' \in \RZ^2 \ .
$$
We shall regard $\CZ^*_1,  \RZ^2$ as 
the subsets of $\GZ(d)$ by the following 
identification of elements, 
$$
 \CZ^*_1 \ni \alpha := (\alpha, 0) \in \GZ(d) \ , \ \ \ 
 \RZ^2 \ni v := (1, v) \in  \GZ(d) \ .
$$
Then $\GZ(d)$ is generated by elements of
$\RZ^2$  with the relations,
$$
v \cdot v' = e^{2 d \pi \sqrt{-1}  <v, v'>} v'
\cdot v =  e^{- 2 d \pi \sqrt{-1}  v_2v_1'} (v+v')
\in \GZ(d) 
\ , \ \ 
 \ {\rm for} \ \ v, \ v' \in 
\RZ^2 . 
$$
As  a character of $\GZ(d)$ is always trivial on
its center, the character-group of $\GZ(d)$
is parametrized by the additive
group $\RZ^2$, of which an element $r$  
corresponds the character $\rho^r$ of $\GZ(d)$ with
$\rho^r(v) := e^{2 d \pi \sqrt{-1} <v, r>}$ for $v
\in \RZ^2 \subset 
\GZ(d)$. 
We shall denote $\Lambda (d)$ the subgroup of $\GZ(d)$ generated by 
integral elements in $\RZ^2$. When $d$ is
an integer, 
$\Lambda (d)$  is the abelian lattice  group
$\ZZ^2$, which  will be denoted  by $\Lambda
$ in what follows.  Associated to an element $
\tau $ of the upper half-plane $ \HZ$, there is a
$\GZ(d) $-representation 
$ {\cal T} (\tau ; d )$ 
 on the space of entire functions of $\CZ$,
$$
{\cal T} ( = {\cal T} (\tau ; d )) : {\cal O}(\CZ) \times \GZ (d) \longrightarrow {\cal O}(\CZ) \ , \ \ 
(\varphi, (\alpha, v) ) \mapsto \varphi | {\cal T}_{(\alpha,v)} : = 
\alpha ( \varphi | {\cal T}_{v} ) \ , 
$$
where   $
(\varphi | {\cal T}_v)(z) : = 
e^{ d \pi \sqrt{-1}( v_1^2 \tau + 2v_1(z+v_2) )} 
\varphi ( z + v_1 \tau + v_2 )$ for $v  \in \RZ^2, 
z \in \CZ
$.  In particular, the following relations
hold,   
$$
{\cal T}_{(v_1,v_2)} = {\cal T}_{(v_1,0)}{\cal
T}_{(0, v_2 )} \ , \ \
 ( \varphi | {\cal T}_{(v_1, 0)}  ) ( z ) =  e^{
d \pi \sqrt{-1}( v_1^2 \tau + 2v_1z )} 
\varphi ( z + v_1 \tau ) \ , \ \ \ \
( \varphi | {\cal T}_{(0,v_2)} )  (z) = 
\varphi (
z + v_2 ) \ . 
$$
For  $ r \in \RZ^2$, the 
multiplication of the 
character $\rho^r$ on ${\cal T}$ gives rise to a 
$\GZ (d)$-representation   $ {\cal T}^r 
( = {\cal T}^r (\tau ; d ) )$ 
on ${\cal O}(\CZ)$, 
$$
(\varphi| {\cal T}^r_v  ) (z) = e^{2  \pi
\sqrt{-1} d <v, r>}  (\varphi| {\cal T}_v) (z) = 
e^{ d \pi \sqrt{-1}( v_1^2 \tau + 2v_1(z+v_2) + 2
<v, r> )} 
\varphi ( z + v_1 \tau + v_2 ) \  \ .
$$
By $ {\cal T}_r {\cal T}^r_* = {\cal T}_* {\cal T}_r$, 
${\cal T}^r$ is equivalent to ${\cal T}$ via
the intertwining operator 
${\cal T}_r$. For 
$d \in \RZ^*, r \in \RZ^2$ and 
$\phi \in {\cal O}(\CZ \times \HZ)$, we shall denote 
\bea(l)
\phi| {\cal T}^r  ( = \phi| {\cal T}^r (d)): = \phi(*, \tau)| 
{\cal T}^r (\tau; d) 
\in {\cal O}(\CZ \times \HZ) \ .
\elea(ellat)
The usual theta function
theory is referred to the case $d=1$. Recall the
theta function $\vartheta ( z , \tau ) $ is an 
element in ${\cal O}(\CZ
\times
\HZ)$ with the   infinite product representation
\[
\vartheta ( z , \tau )   =   \prod_{n=1}^{ \infty} (
1 + 2 q^{n-\frac{1}{2}} \cos 2 \pi {\rm z} + q^{2n-1} ) 
( 1 - q^{n} ) \ , \ \ q = e^{2\pi \sqrt{-1}\tau}
\ , \ \
\mbox{Im} (\tau)
\gg 0
\ ,
\]
while the theta function with characteristics 
$r=(a,b) \in \RZ^2$ is defined by
\footnote{Here we write the theta function $\vartheta^{(a, b)} $ 
instead of  $\vartheta [\begin{array}{c} a \\ b 
\end{array} ]$ in usual literature.}
$$
\vartheta^r ( z , \tau ) : = 
e^{  \pi \sqrt{-1} ( a^2 \tau + 2 a ( z + b )) }
\vartheta( z + a \tau + b , \tau )  , \ \ ( z , \tau ) \in {\CZ} \times \HZ 
\ .
$$ 
For a fixed $\tau$, $\vartheta ( z, \tau ) $ (
or  $\vartheta^r ( z, \tau )$ ) is the only     
$\Lambda$-invariant entire function on 
$\CZ$, up to a constant multiple, for the
action  ${\cal T}(1)$ ( 
${\cal T}^r(1)$ resp.) . In fact, one has the 
following expression of the theta
functions,
$$
\begin{array}{lll}
 \vartheta ( z , \tau ) & = 
( {\bf 1} |\sum_{m = - \infty }^{\infty }   
{\cal T}_{(m,0)} (1)  ) ( z , \tau  )  
&= \sum_{m= - \infty }^{
\infty }  e^{  \pi i (m^2 \tau +2m z) } \ , \\
\vartheta^r ( z , \tau )    
 &= ( ( {\bf 1} | {\cal T}^r_r(1))|\sum_{m = - \infty }^{\infty }
 {\cal T}^r_{(m,0)} (1)) ( z , \tau ) 
&=  \sum_{m= - \infty }^{
\infty }e^{ \pi \sqrt{-1}( (m+a)^2 \tau +
2(m+a)(z + b )) }
\  ,
\end{array}
$$ 
where ${\bf 1}$ is the constant function
with value one. Associated to a theta function,
one has the theta form,
$$
\Theta (z, \tau)  = \frac{ \vartheta (z, \tau)}{\eta (\tau)^3 } \ , \ 
\ \ 
\Theta^r  (z, \tau)  = \frac{ \vartheta^r (z, \tau)}{\eta (\tau)^3 } \ 
\in {\cal O}(\CZ \times \HZ) \ ,  
$$
where $\eta ( \tau) \ ( = q^{1/24}
\prod_{n=1}^{\infty}(1-q^n) )$  is the Dedekind
eta function. The following relation holds,
$$
\Theta^{(a,b)} ( z , \tau )  = 
e^{  \pi \sqrt{-1} ( a^2 \tau + 2 a ( z + \mu )) }
\Theta ( z + a \tau + b , \tau )  , \ \ 
( z , \tau) \in {\CZ} \times \HZ \ .
$$
Corresponding to Jacobi theta function  
$\vartheta_1  ( z, \tau ) \ (: =  
\vartheta^{( \frac{1}{2}, \frac{1}{2})} ( z , \tau ))$, 
we have the Jacobi-form 
\bea(l) 
\Theta_1(z, \tau) := \Theta^{( \frac{1}{2}, \frac{1}{2})}  
(z, \tau) \ . 
\elea(Psi1)
It is known that the asymptotic expansion of 
$\Theta_1(z, \tau)$ near $z=0$ is given by   
$$  
 \Theta_1(z, \tau)  = 2 \pi z + O(z^2) \ \ \ \ \ 
{\rm as} \ z \longrightarrow 0 \ . 
$$
We have the
quasi-periodicity  and zero relations for
$\Theta^{(a,b)} ( z , \tau )$,
\bea(l)
\Theta^{(a, b)} ( z + 1 , \tau )  =  e^{ 2 \pi {\rm i } a}
\Theta^{(a, b)} ( z , \tau ) ,  \ \ \ 
\Theta^{(a, b)} ( z +  \tau , \tau )  =  e^{ - \pi
\sqrt{-1} ( \tau + 2 ( z + b)) } 
\Theta^{(a, b)} ( z , \tau ) ,   \\
\Theta^{(a, b)} ( z  , \tau ) = 0  \ \ \  \Longleftrightarrow \ \ \ 
z \equiv ( \frac{1}{2} - a) \tau + ( \frac{1}{2} - b ) \
\ \ ( \mbox{mod.} \ \ {\ZZ} \tau + {\ZZ} ) \ .
\elea(pzPsi)
One has 
\bea(l)
\Theta^{(a + 1, b)}(z, \tau) = \Theta^{(a , b)}(z, \tau) \ , \ \ \ \  
\Theta^{(a , b+1 )}(z, \tau) = e^{2 \pi
\sqrt{-1}a} 
\Theta^{(a , b)}(z, \tau) \ , \\
\Theta^{(a , b)}(- z, \tau) = e^{2\pi \sqrt{-1}
a} 
\Theta^{(1-a ,1- b)}(z, \tau) \ , \\
\Theta^{(a + a' , b + b')}(z, \tau) =e^{\pi
\sqrt{-1}(a^2 \tau +  2a(z+b+b'))} 
\Theta^{(a' , b')}(z + a \tau + b , \tau) \ .
\elea(threl)
By lifting of the geometrical modular
action, we have  a $SL_2(\RZ)$-representation  
 on ${\cal O}(\CZ \times \HZ)$ for each pair 
$(w, d)$ with
$ w \in \frac{1}{2}\ZZ, d \in \RZ^*$,  
\bea(ll)
\Upsilon (= \Upsilon ( w, d) ) :& {\cal O}(\CZ \times \HZ) \times SL_2(\RZ) \longrightarrow 
{\cal O}(\CZ \times \HZ) \ , \ \ \ \ (\phi, M) \mapsto \phi|\Upsilon_M \ ,
\\
& (\phi|\Upsilon_M) (z, \tau) : = (C\tau+D)^{-w}
e^{ \frac{-d \pi \sqrt{-1}Cz^2 }{C\tau+D}} 
\phi(\frac{z}{C\tau+D}, \frac{A\tau +B}{C\tau+D}) \ \ . 
\elea(modat)
The relation of the 
elliptic and modular 
actions on ${\cal O}(\CZ \times \HZ)$ is given
by the following lemma.
\par \vspace{.2in} \noindent
{\bf Lemma 1.} For $(w, d) \in
\frac{1}{2}\ZZ \times \RZ^*$,  we denote ${\cal
T}^r= {\cal T}^r(d),
\Upsilon = \Upsilon ( w, d)$.  Then the following
relation holds  for $M \in
SL_2(\RZ)$, $v, v' \in \RZ^2$ with 
$v=v'M$, and ${\rm for} \ \phi \in 
{\cal O}(\CZ \times \HZ)$:
$$ 
e^{  - d \pi \sqrt{-1}(v_1v_2 + 2<v, r>)} 
(\phi|\Upsilon_M) | {\cal T}^r_{v} =
 e^{ - d \pi \sqrt{-1}(v_1'v_2' + 2<v', r>)} 
(\phi|{\cal T}^r_{v'})|\Upsilon_M \ \ .
$$
\par \vspace{.1in} \noindent 
{\it Proof.} By the definition of ${\cal T}^r$,
one needs only to consider the case $r=0$. Note
that $v_1= v_1'A+v_2'C,  v_2 = v_1'B+v_2'D$. By 
computation, one has the following relations:
$$
\begin{array}{rl}
(C\tau+D)^{w}((\phi|{\cal T}_{v'})|\Upsilon_M) 
(z, \tau)  =& 
e^{ d \pi \sqrt{-1}\frac{-Cz^2+2v_1'z +
(v_1'^2 A+2v_1'v_2'C )\tau +v_1'^2B+2v_1'v_2'D
}{C\tau+D}} 
\phi ( \frac{z+ v_1\tau +v_2}{C\tau+D} , 
\frac{A\tau +B}{C\tau+D}) \ , \\
(C\tau+D)^{w}((\phi|\Upsilon_M) | {\cal T}_{v})(z, \tau)  
=&  e^{ d \pi \sqrt{-1}( 
 \frac{-Cz^2  + 2v_1'z + 
 (v_1'^2A+2v_1'v_2'C)\tau+ v_1'^2B+2v_1'v_2'D 
 }{C\tau+D}+v_1'^2AB+2v_1'v_2'BC+v_2'^2CD)} \\
&\phi(\frac{z + v_1 \tau + v_2}{C\tau+D},
\frac{A\tau +B}{C\tau+D}) \ .
\end{array} 
$$
By 
$v_1v_2-v_1'v_2' = 
v_1'^2AB+2v_1'v_2'BC+v_2'^2CD$, the result
follows immediately.
$\Box$ \par \vspace{.2in} \noindent
The conclusion of the above lemma has indicated
that one can not lift the geometric action of 
$\RZ^2 * SL_2(\RZ)$ of
$\CZ \times \HZ$ directly 
to one of  
${\cal O}(\CZ \times \HZ)$. However, the
formulas (\req(ellat)),(\req(modat)) do give
rise to a module structure of ${\cal O}(\CZ
\times
\HZ)$ for the semi-product $\GZ(d) \stackrel{r}{*}
SL_2(\RZ)$ of 
$\GZ(d)$ and $SL_2(\RZ)$.  Here in the group
$\GZ(d)
\stackrel{r}{*} SL_2(\RZ)$,
 the conjugate relation 
of $M \in SL_2(\RZ)$  on $v \in \RZ^2
\subseteq 
\GZ(d) $ is defined by
\bea(l)
M \cdot v \cdot M^{-1} = e^{  d \pi \sqrt{-1}
(v_1v_2-v_1'v_2' + 2<v-v', r>)} v' 
 ( = e^{  d \pi
\sqrt{-1}((v_1-2a)(v_2+2b)-(v_1'-2a)(v_2'+2b))}
v'  ) 
\ , 
\elea(HMgp)
where $v'=(v_1', v_2')$ is the
(matrix-product) element 
$v M^{-1}$. Note that
$\GZ(d) \stackrel{r}{*} SL_2(\RZ)$ is a
central extension  of $\RZ^2 * SL_2(\RZ)$ with
its center contained in
$\GZ(d)$. Hence we have obtained the following
result:   
\par \vspace{.2in} \noindent
{\bf Proposition 1.} The relations
(\req(ellat)), (\req(modat))  give rise to the
representation of $\GZ(d) \stackrel{r}{*}
SL_2(\RZ)$ on 
${\cal O}(\CZ \times \HZ)$
\bea(ll)
 {\cal J}^{d, r, w} &: {\cal O}(\CZ
\times \HZ) \times  ( \GZ(d) \stackrel{r}{*}
SL_2(\RZ) ) 
\longrightarrow {\cal O}(\CZ \times \HZ) .
\elea(HeisJ)
$\Box$ \par \vspace{.2in} \noindent
For the rest of this section, we shall only
consider the  case  $d \in \ZZ$, in which situation
one will see that the representation (\req(HeisJ))
does descend to a certain arithmetic subgroup of
$\RZ^2 * SL_2(\RZ)$.  Denote $\Gamma$ the
full modular group, $\Gamma = SL_2 ( \ZZ ) $. 
The subgroup 
$\Gamma^{\rm J} := \Lambda * \Gamma$ of $\RZ^2
* SL_2(\RZ)$ is called   the Jacobi group. In
general,   the Jacobi group associated to a
subgroup $\Gamma'$ of $\Gamma$ is defined  by
$\Gamma'^{\rm J} : =
\Lambda' * \Gamma$. It is easy to see that 
the following elements generates
$\Gamma^{\rm J}$, 
$$
{\rm u} = (1, 0) , \ {\rm v}= (0,1) \ \in \ZZ^2
\ , \ \ \  {\rm T} =  \left( \begin{array}{cc}
1&1\\ 
0&1
\end{array}
\right)  , \  {\rm S } =  \left( \begin{array}{cc}
0&1\\ 
-1&0
\end{array}
\right)  \in \Gamma .
$$ 
In fact, $\Gamma^{\rm J}$ is
characterized as the group  generated by four 
elements ${\rm u, v}$, T, S, with the relations,
$$
\begin{array}{lll}
{\rm u} {\rm v} = {\rm v} {\rm u}\ ,  &
 {\rm S}^4 = 1 \ , & \ 
{\rm S}^2 {\rm T} = {\rm T}{\rm S}^2 \ , \ \\ 
{\rm T}^{-1}{\rm u}{\rm T}={\rm u} + {\rm v} \ ,
&   {\rm T}^{-1} {\rm v} {\rm T} = {\rm v} \
, & {\rm S}^{-1} {\rm u} {\rm S} = {\rm v} \
, \ \ {\rm S}^{-1} {\rm v} {\rm S} = {\rm
u}^{-1} \ .
\end{array}
$$
One can easily see that    
a character $\lambda$ of $\Gamma^{\rm J}$ is always
trivial  on the subgroup $\Lambda$, hence
determined   by the values of 
$\lambda ({\rm T}), \lambda ({\rm S})$.
Besides the full modular group $\Gamma$, the
modular groups we shall
also concern in this
paper are the following one for $q = (q_1, q_2) 
\in \QZ^2$,
$$
\Gamma_q : 
= \{ M \in \Gamma \ | \  q M - q \in \Lambda  \} 
\ (= 
\{ M \in \Gamma \ | \ 
 e^{2 \pi \sqrt{-1} < vM-v , q >} = 1  \  
\forall \  v \in \Lambda   \} ) 
\ \ .
$$
Note that $\Gamma_q$ depends only on the class of $q$ in 
$\QZ^2/\Lambda$. The modular group 
$\Gamma_{(\frac{1}{2},
\frac{1}{2})}$ is  generated by 
 S , ${\rm T}^2$, which  was denoted
by
$\Gamma_{\theta}$ in \cite{R}. 
\par \vspace{.2in} \noindent 
{\bf Proposition 2. } Associated to an element $ r
$ of $\QZ^2$, we denote   
$q= (\frac{1}{2}, \frac{1}{2})-r$.
Then the following relation holds 
for $d, 2w \in \ZZ, v \in \Lambda$ and $ M \in
\Gamma_q$,
$$
 (\phi | \Upsilon_M(w, d))|{\cal T}^r_{vM}(d) = 
(\phi | {\cal T}^r_v(d))|\Upsilon_M(w, d)  \ , \ \ 
\ \ \ \ \phi \in  {\cal O}(\CZ \times \HZ)
\ . 
$$
\par \vspace{.1in} \noindent
{\it Proof.} Since the sign
$(-1)^{(m+1)(n+1)}$ in unchanged under the action
of
$\Gamma$  for  $ (m,n) \in \Lambda$, the above
equality holds  when $r=
(\frac{1}{2},
\frac{1}{2}) $.  Then it follows the result for an
arbitrary $r$ by the definition of 
$\Gamma_q$.
$\Box$ \par \vspace{.2in} \noindent
{\bf Remark.} Applying the same argument in the
proof of the above proposition  to the
representation (\req(HeisJ)) for 
$d \in
\QZ$, one obtains a
$\Lambda(d) 
\stackrel{r}{*} 
\Gamma_q$-module structure  of 
${\cal O}(\CZ \times \HZ)$.
$\Box$ \par \vspace{.1in} \noindent
For simplicity, we shall write $\phi |v, \phi
|M
$ instead of $\phi  | {\cal T}^r_v(d) , \phi |
\Upsilon_M(w, d)$   in the case  
$d \in \ZZ$ if no confusion 
could arise. By 
Proposition 2, one obtains an 
$\Gamma^{{\rm J}_q}$-representation on
${\cal O}(\CZ \times \HZ)$, called 
the weight $w$ and index $\frac{d}{2}$ 
representation of $\Gamma^{{\rm J}_q}$,
\bea(cl)
{\cal O}(\CZ \times \HZ) \times \Gamma_q^{\rm J} \longrightarrow 
{\cal O}(\CZ \times \HZ) \ ,  & ( \phi , *) \mapsto \phi|* \ ,
\\
( \phi| v )(z, \tau) = e^{ d \pi \sqrt{-1} 
(m^2 \tau + 2  
 m z +2m b - 2n a)} \phi( z + m \tau + n, \tau) \ , & v = (m,n) 
\in \Lambda \ , \\
(\phi|M) (z, \tau) = (C\tau+D)^{-w}
e^{ \frac{-d\pi \sqrt{-1}Cz^2 }{C\tau+D}} 
\phi(\frac{z}{C\tau+D}, \frac{A\tau +B}{C\tau+D}) \ , 
& M \in \Gamma_q, 
\elea(GJrep)
where $ q= (\frac{1}{2}-a, \frac{1}{2}-b)$. 
In particular, 
for $a=b= \frac{-1}{2}$ (or $\frac{1}{2}$), this is
the usual weight $w$ and index
$\frac{d}{2}$ $\Gamma^{\rm J}$-representation of
which the invariant function $\phi (z, \tau)$  is
called a ($\Gamma^{\rm J}$-)Jacobi form  of 
weight $w$ and index $\frac{d}{2}$, i.e., the
function 
$\phi (z, \tau)$ satisfies the following
elliptic and modular properties \cite{EZ}:
$$
\begin{array}{ll}
\phi( z + m \tau + n, \tau) 
 =  e^{ -d \pi \sqrt{-1} (m^2 \tau + 2  
 m z -m +n) } \phi(z, \tau)\ , & (m,n) \in \Lambda
\ , \\
\phi(\frac{z}{C\tau+D}, \frac{A\tau +B}{C\tau+D})
 = (C\tau+D)^{w}
e^{ \frac{d\pi \sqrt{-1}C z^2 }{C\tau+D}} 
\phi(z, \tau) \ , 
& M \in \Gamma \ .
\end{array}
$$
 Denote $
{\cal O}(\CZ \times \HZ)_{w,\iota}$ the 
space of all Jacobi forms with weight $w$ and
index $\iota$. 
Then with the canonical grading, the Jacobi forms
form a graded algebras,
$\bigoplus_{w, \iota \in \frac{1}{2} \ZZ } 
{\cal O}(\CZ \times \HZ )_{w,\iota} 
$.
An important example of Jacobi form 
is the  Jacobi-theta form $\Theta_1 (z, \tau)$
defined in (\req(Psi1)), in fact, we have
\bea(l)
\Theta_1(z, \tau)  \in 
{\cal O}(\CZ \times \HZ)_{ -1 , \frac{1}{2}} \ ,
\elea(Psi1m)
a conclusion followed by the facts that
$\eta ( \tau)$ is a weight
$\frac{1}{2} \ \Gamma$-modular eigenform  for the
eigencharacter 
$\sigma$, and $\vartheta_1(z, \tau)$ is
a  Jacobi eigenform of weight 
$\frac{1}{2}$ and index $\frac{1}{2}$ with the 
eigencharacter 
$\sigma^3$, where 
$\sigma$ is the
character of $\Gamma$  defined by $\sigma ( {\rm
T}) = e^{\frac{\pi \sqrt{-1}}{12}} , 
\sigma ( {\rm S}) = e^{\frac{\pi \sqrt{-1}}{4}}$.
\par \noindent
{\bf Remark.} The discussion of this section on
the Jacobi-form theory can be naturally extended to
meromorphic forms on $\CZ \times \HZ$.

\section{N=2 Superconformal Algebra}
In this section, we shall derive the theta-form
expression of characters of N=2 superconformal
algebra representations.  First we recall the some
well-known facts on the 
algebra. 

{\bf Definition.} The
N=2 superconformal algebra,  denoted by ${\sf
SCA}$,   consists of the stress tensor
$L(\zeta),$   two super-currents
$G^{\pm}(\zeta)$ and a $U(1)$-current
$J(\zeta)$, 
$$
L (\zeta) = \sum_{n \in \ZZ} L_n \zeta^{-n-2} \ , \ \ \ 
G^{\pm}(\zeta) = \sum_{p \in \frac{1}{2} + \ZZ} 
G^{\pm}_p \zeta^{-p - \frac{3}{2}} \ , \ \ \ 
J(\zeta) = \sum_{n \in \ZZ} J_n \zeta^{-n-1} \ , 
$$
where the coefficients, $L_m, J_n, G^{\pm}_p$, 
form a super-Lie algebra with a central element $c$ : 
$$
\begin{array}{ll}
[ L_m , L_n ] = (m-n) L_{m+n} + \frac{c(m^3-m)}{12} \delta_{m+n,0} \ , &
[ J_m , J_n ] = \frac{c m}{3}  \delta_{m+n, 0} \ , \ \ \ \  
[L_m , J_n ] = - n J_{m+n} \ , \\

[ L_m, G_p^{\pm} ] = ( \frac{m}{2}- p ) G_{m+p}^{\pm} \ , & 
[ J_m, G_p^{\pm} ] = \pm G_{m+p}^{\pm} \ , \\
\{ G_p^+, G_q^+ \} = \{ G_p^-, G_q^- \} = 0 \ , & 
\{ G_p^+, G_q^- \} = 2 L_{p+q} + ( p-q)J_{p+q} + 
\frac{c}{3}(p^2 - \frac{1}{4}) \delta_{p+q, 0}  
\ .
\end{array}
$$
We shall not distinguish the central element $c$ 
with its eigenvalues. For convenience,  we shall use the bold letter
$\bf c$ to  denote one-third of the central
element, 
$$    
{\bf c} := \frac{c}{3} \ . 
$$
There exists the
$\ZZ$-symmetry of the algebra  
${\sf SCA}$, called the spectral flow, which is generated by 
the automorphism,  
$$
L_n \mapsto L_n +  J_n + \frac{\bf c}{2}  
\delta_{n,0} \ , \ \ J_n \mapsto J_n + {\bf
c}  
\delta_{n,0} \ , \ \ 
G^+_p \mapsto G^+_{p + 1} \ , \ \
G^-_p \mapsto G^-_{p - 1} \ .
$$
In the topological field theory, it is more convenient to consider 
another set of generators of ${\sf SCA}$ associated to 
a real parameter $\frac{-1}{2} \leq a \leq
\frac{1}{2}$ by the following twisting currents: 
$$
\begin{array}{ll}
L^a (\zeta) = L (\zeta) + a \partial J (\zeta) + 
\frac{{\bf c} a^2 }{2}\zeta^{-2} = 
\sum_{n \in \ZZ} L^{a}_n \zeta^{-n-2} \ , &
J^{a}(\zeta) = J (\zeta) + {\bf c} \zeta^{-1}= 
\sum_{n \in \ZZ} J^{a}_n \zeta^{-n-1}  \ , \\
G^{a}(\zeta) = G^+(\zeta) = \sum_{p \in - a + \frac{1}{2} + \ZZ}
G^{a}_p \zeta^{-p-a-\frac{3}{2}} \ , &
Q^{a}(\zeta) = G^-(\zeta) = \sum_{p \in a + \frac{1}{2} + \ZZ} 
Q^{a}_p \zeta^{-p+a  -\frac{3}{2}} \ ,
\end{array}
$$
with the coefficients 
$$
L^{a}_n = L_n +a (n+1) J_n + \frac{ {\bf c} a^2
}{2} 
\delta_{n,0} \ , \ \ \ 
J^{a}_n = J_n + {\bf c} a  \delta_{n,0}\ , \
\  G^{a}_p = G^+_{p+ a } \ , \ \ \ 
Q^{a}_p = G^-_{p -a } \ , 
$$
satisfying the relations,
$$
\begin{array}{ll}
[ L^{a}_m , L^{a}_n ] = (m-n) L^{a}_{m+n} + 
 \frac{{\bf c}(m^3(1-4a^2)-m)}{4} 
\delta_{m+n,0}  ,& 
 [ J^{a}_m , J^{a}_n ] = {\bf c}m 
\delta_{m+n,0}  , 
\\

[ L^{a}_m , J^{a}_n ] = - n 
J^{a}_{m+n} +  {\bf c} n^2  
\delta_{m+n,0} , &  \\

[ L^{a}_m, G^a_p ] = (\frac{(1+2a)m}{2} -p) 
G^{a}_{m+p} , & 
[ L^{a}_m, Q^a_p ] = (\frac{(1-2a)m}{2} -p)
Q^{a}_{m+p} ,  \\ 

[ J^{a}_m , G^{a}_p ] = G^{a}_{m+p} , \ \ \ \ \ 
[ J^{a}_m , Q^{a}_p ] = - Q^{a}_{m+p} , &
\{ G^{a}_p , G^{a}_q \} =  
\{ Q^{a}_p , Q^{a}_q \} = 0 
,  \\ 
\{ G^{a}_p, Q^{a}_q \} = 2 
L^{a}_{p+q}+((1-2a) p-(1+2a)q) J^{a}_{p+q} + 
\frac{{\bf c}(4p^2-1)}{4} 
\delta_{p+q,0} \ .
\end{array}
$$
One has the  decomposition, $ {\sf SCA} = {\sf
SCA}_+
\oplus {\sf SCA}_0 \oplus  {\sf SCA}_- $, 
where
$$
\begin{array}{lll}
{\sf SCA}_0 = &  <c, L_0, J_0> & = \ <c, L^a_0, J^a_0>  \ , \\
{\sf SCA}_+ = & <L_m , J_m ,  G^\pm_p \ | \ 
 m> 0 , p \geq 0 >  & = \ 
<L^a_m , J^a_m ,  G^a_p , Q_p^a \ | \  m> 0 , p \geq 0 > \ , \\
{\sf SCA}_- = & < L_m , J_m ,  G^\pm_p \ | \ 
 m< 0 , p \leq 0 >  & = \ 
<L^a_m , J^a_m ,  G^a_p , Q_p^a \ | \  m< 0 , p \leq 0 > \ .
\end{array}
$$ 
A highest weight module (HWM)  is a 
${\sf SCA}$-module generated by a highest vector, 
i.e., 
an eigenvector of ${\sf SCA}_0$ and annihilated by 
${\sf SCA}_+$.  The letters $ H, Q$ will denote 
the $ L_0, J_0$-eigenvalues  of the highest vector
 respectively;    
similarly the letters $H^a, Q^a$ for $L^a_0,
J^a_0$-eigenvalues. Then  
$H^a = H + a Q + \frac{a^2{\bf c}}{2},  Q^a = Q + a {\bf c} $.
The character of a HWM is the following Laurent
series,
$$
{\rm NS} (z, \tau) = {\rm Tr} (q^{L_0 -\frac{{\bf c}}{8}} y^{J_0} ) \ 
 \ \ , \ \ y := e^{2\pi \sqrt{-1} z},  
q := e^{2\pi \sqrt{-1} \tau} \ .
$$
The twisted character with characteristic $(a,
b)$  for $\frac{-1}{2} \leq a, b \leq
\frac{1}{2}$ is defined by 
\bea(l)
{\rm Ch}^{(a,b)} (z, \tau) = {\rm Tr} (q^{L_0^a-\frac{{\bf c}}
{8}} (e^{2\pi \sqrt{-1} b}y)^{J_0^a} ) = 
e^{{\bf c} \pi 
\sqrt{-1}(a^2 \tau + 2 a (z+ b ))}
{\rm NS} (z+ a \tau + b, \tau) 
\elea(Ch)
By 
$$
\begin{array}{llll}
L_0^{1-a}=& L_0^a + (1-2a)J_0^a + \frac{(1-2a)^2 
\bf c}{2},&
J_0^{1-a}=& J_0^a + (1-2a){\bf c}  ; \\ 
L_0^{\frac{1}{2}} =& L_0^{\frac{-1}{2}} +
J_0^{\frac{-1}{2}} + 
\frac{\bf c}{2}
 , & 
J_0^{\frac{1}{2}} =& J_0^{\frac{-1}{2}} +{\bf c}
, 
\end{array}
$$
one obtains the following relations of twisted characters,
\bea(l)
{\rm Ch}^{(1-a, b)} (z, \tau) = 
e^{{\bf c}\pi \sqrt{-1}((1-2a)^2 \tau + 2 
(1-2a)(z+b)) }{\rm Ch}^{(a, b)}(z + (1-2a)\tau , \tau) , \\  
{\rm Ch}^{(a, 1- b)} (z + (1-2b), \tau)=  
{\rm Ch}^{(a, b)} (z, \tau) \  , \\
{\rm Ch}^{(\frac{1}{2}, b)} (z, \tau) = e^{ {\bf c}\pi
\sqrt{-1}  (\tau + 2(z+b) }
{\rm Ch}^{(\frac{-1}{2}, b)}(z+\tau , \tau) \ , \ \ \ \ 
{\rm Ch}^{(a, \frac{1}{2})} (z, \tau)=  
{\rm Ch}^{(a, \frac{-1}{2})} (z +1, \tau) \ .
\elea(chs)
We shall denote\footnote{The letters, NS, R, indicate the  
Neveu-Schwarz or Ramond sectors. } 
$$
{\rm R} (z, \tau) =  {\rm Ch}^{(\frac{1}{2},\frac{1}{2})} (z, \tau) \ , \ \ \ 
\widetilde{\rm R} (z, \tau) = 
{\rm Ch}^{(\frac{-1}{2},\frac{-1}{2})} (z, \tau)
\ .
$$
For $0<{\bf c}<1$, 
all the unitary irreducible HWM of ${\sf SCA}$
are labelled by three integers  
$k, m, l $ with the relations, $  
|m| \leq l \leq k \ , \ l \equiv m \pmod{2}$, where ${\bf c}, H, Q$ 
are given by
\bea(l)
{\bf c}= \frac{k}{k+2} \ , \ \
H (= H_{l,m} ) = \frac{l^2+2l-m^2}{4(k+2)} \ , \ Q (= Q_{l,m} )
= \frac{m}{k+2} .  
\elea(DS)
The character of the corresponding HWM will be
denoted by $ {\rm Ch}_{l,m}^{(a,b)} (z, \tau) $. 
A HWM is called chiral (resp. antichiral) 
if $H_{l,m} = \frac{Q_{l,m}}{2}$ ( resp. 
 $ \frac{- Q_{l,m}}{2}$ ),  
characterized by the highest vector annihilated
by 
$G^+_{\frac{-1}{2}}$ (resp. $G^-_{\frac{-1}{2}}$ ) .
For the rest of this paper, associated to a
positive integer $k$ we shall use the bold letter 
$\bf k$ to denote the number $k+2$, 
$$
{\bf k} : = k+2 \ . 
$$
To a pair of integers $l,
m$ with 
$l \equiv m \pmod{2}$, we shall always use 
the letters $i, j$ to denote 
the following half-integers : 
$$
 i= \frac{l-m+1}{2} \ , \ \ \ j = 
\frac{-(l+m+1)}{2}
\ \ \in 
\frac{1}{2} + \ZZ .
$$
With a fixed central element ${\bf c} =
 \frac{{\bf k}-2 }{\bf k}$,  the unitary
irreducible HWMs 
are indexed by the following equivalent data,
$$
\begin{array}{lll}
(l, m) \in \ZZ^2 \ ,  \ |m| \leq l \leq k \ ,  \ l \equiv m \pmod{2} 
& \Longleftrightarrow & (i, j ) \in (\frac{1}{2} + \ZZ)^2 \ , \ 
0 < i, -j  , i - j < {\bf k} \ ,
\end{array}
$$
hence the chirality is given by  
$$
\begin{array}{llll}
{\rm chiral \ HWM } : & m = l , &
\Longleftrightarrow &  i = \frac{1}{2} \ , \\
{\rm antichiral \ HWM } : & 
m = -l , & \Longleftrightarrow & j = \frac{-1}{2} \ . \\
\end{array}
$$
Then (\req(DS)) becomes 
\bea(l)
H =  \frac{-4ij-1}{4{\bf k}} \ , \ \ 
H^{a}= 
\frac{l^2+2l -(m-2a)^2}{4{\bf k}} + 
\frac{a^2}{2} = \frac{-4(i+a)(j+a)-1}{4{\bf k}} + \frac{a^2}{2} \ , \\
Q = \frac{-(i+j)}{\bf k} \ , \ \ Q^{a} = \frac{m + a ({\bf k}-2)}{\bf k} 
= \frac{-(i+j) + a ({\bf k}-2)}{\bf k} \ .
\elea(chij)
For the rest of this section, we are going to
derive the
theta-form  expression of the character of the
above HWM and discuss the functional   properties
among these characters.  First  let us
recall the Laurent series expression
\footnote{ Here we adhere to papers 
\cite{D, M, RY} for the mostly suitable
format to  our treatment.}
of ${\rm NS}_{l,m}(z, \tau)$,
$$
{\rm NS}_{l,m}(z, \tau) = q^{H-\frac{\bf c}{8}} y^Q  
\gamma_{l,m}(z, \tau) \varphi (z , \tau) \ ,
$$
where 
$$
\begin{array}{ll}
\varphi (z, \tau) &=  
\frac{ \prod_{n = 1}^{\infty}(1+ y q^{n - 1/2})(1+ y^{-1} q^{n - 1/2})}
{\prod_{n=1}^{\infty}(1-q^n)^2} , \\
\gamma_{l,m}(z, \tau) &= \frac{ 
\prod_{n = 1}^{\infty}(1-q^{{\bf k}(n-1)+i-j})(1-q^{{\bf k}n-i+j})
(1-q^{{\bf k}n})^2}{\prod_{n=1}^{\infty}
(1+yq^{{\bf k}(n-1)+i}) (1+y^{-1}q^{{\bf k}n-i})
(1+yq^{{\bf k}n+j})(1+y^{-1}q^{{\bf k}(n-1)-j})} \ .
\end{array}
$$
\par \vspace{0.1in} \noindent 
{\bf Theorem 1.} The characters 
${\rm NS}_{l,m}(z, \tau) , 
{\rm Ch}_{l,m}^{(a, b)}(z, \tau)$ are elements 
in ${\cal O}(\CZ \times \HZ)$ with the following 
theta-form representation:
\bea(ll)
{\rm NS}_{l,m}(z, \tau) & = 
\frac{  
e^{ \pi\sqrt{-1}(\frac{1}{2}- \frac{i-j}{{\bf
k}})} 
 \Theta^{(
\frac{1}{2} + \frac{i-j}{{\bf k}}, \frac{1}{2})
}(0, {\bf k} \tau) \Theta (z, \tau) }
{
\Theta^{ (\frac{1}{2} +\frac{i}{{\bf k}}, 0)} ( z  , {\bf k} \tau) 
\Theta^{ ( \frac{1}{2} + \frac{j}{{\bf k}}, 0)} ( z , {\bf k} \tau) } \ , 
\\
{\rm Ch}_{l,m}^{(a, b)}(z, \tau) &= \frac{ 
e^{ \pi\sqrt{-1}(\frac{1}{2}- \frac{i-j}{{\bf
k}})}
 \Theta^{
(\frac{1}{2} + \frac{i-j}{{\bf k}} , \frac{1}{2})} (0, {\bf k} \tau) 
\Theta^{(a, b)} (z, \tau) }
{
\Theta^{( \frac{1}{2} +\frac{i+a}{{\bf k}}, b )}
 ( z  , {\bf k} \tau) 
\Theta^{(\frac{1}{2} + \frac{j+ a}{{\bf k}}, b )} 
( z , {\bf k} \tau) } \ .
\elea(chtheta)
{\it Proof.}   
By the infinite product formula of $\vartheta (z, \tau)$, the 
factors $\varphi(z, \tau), \gamma_{l,m}(z, \tau)
$ in the formula of
 ${\rm NS} (z, \tau)$ have the following expressions:
$$
\varphi(z, \tau)  = q^{\frac{1}{8}}
\Theta (z, \tau) , \ \ \
\gamma_{l,m}(z, \tau) 
= \frac{ q^{\frac{-{\bf k}}{8}}  \Theta ( 
(\frac{1}{2}-\frac{i-j}{\bf k}) \bf k \tau +
\frac{1}{2}, {\bf k}\tau) }
{\Theta (  z + (\frac{-1}{2}+ \frac{i}{\bf k}) \bf k \tau  , {\bf k}\tau)
\Theta ( z + (\frac{1}{2}+\frac{j}{\bf k}) \bf k \tau , {\bf k}\tau) } .  
$$
By (\req(chij)) (\req(threl)), we have 
$$
\gamma_{l,m}(z, \tau) 
= \frac{ q^{-H + \frac{{\bf c}-1}{8}} y^{-Q} 
e^{\pi \sqrt{-1}(\frac{-1}{2} + \frac{i-j}{\bf
k})} 
\Theta^{ ( \frac{1}{2}-\frac{i-j}{\bf k} ,
\frac{1}{2})} (0 , {\bf k}\tau) }
{\Theta^{( \frac{-1}{2}+ \frac{i}{\bf k}, 0) }  (z, {\bf k}\tau)
\Theta^{ (\frac{1}{2}+\frac{j}{\bf k} ,0)} (z, {\bf k}\tau) } = 
\frac{ q^{-H + \frac{{\bf c}-1}{8}} y^{-Q} 
e^{\pi \sqrt{-1}(\frac{1}{2} - \frac{i-j}{\bf
k})} 
\Theta^{ ( \frac{1}{2}+\frac{i-j}{\bf k} ,
\frac{1}{2})} (0 , {\bf k}\tau) }
{\Theta^{( \frac{1}{2}+ \frac{i}{\bf k}, 0) }  (z, {\bf k}\tau)
\Theta^{ (\frac{1}{2}+\frac{j}{\bf k} ,0)} (z, {\bf k}\tau) } \ .
$$
Hence we obtain the expression of ${\rm
NS}_{l,m}(z,
\tau)$. By the last relation of 
(\req(threl)), it follows the expression of  
${\rm Ch}_{l,m}^{(a, b)}(z, \tau)$. The zeros of
the denominator in the 
 expression of ${\rm Ch}_{l,m}^{(a, b)}(z, \tau)$ are given by 
$$
z \equiv -(i+a)\tau + \frac{1}{2}-b \ , \ \  \ \
 -(j+a)\tau + \frac{1}{2}-b \ \ 
\ ( \rm{mod.} \ \ZZ {\bf k}\tau + \ZZ ) \ .
$$
By $0< i-j <{\bf k}$, there is no common solution
for the above two   equations. Hence the
poles of 
${\rm Ch}_{l,m}^{(a, b)}(z, \tau)$ are 
all simple; so are the zeros of
$\Theta^{(a, b)} (z, \tau)$ in  the numerator of
the expression. Then one can easily see that   
${\rm Ch}_{l,m}^{(a, b)}(z, \tau) \in  {\cal O} (\CZ \times \HZ)$. 
$\Box$ \par \vspace{.2in} \noindent
By the last relation of (\req(threl)) for $(a, b) =
\pm (\frac{1}{2}, 
\frac{1}{2})$, the Ramon characters are expressed
by
\bea(ll)
{\rm R}_{l,m} (z, \tau) &= \frac{ \sqrt{-1}
q^{H+\frac{Q}{2}} (-y)^{Q + 
\frac{{\bf c}-1}{2}}
 \Theta_1 ((i-j)\tau, {\bf k} \tau) 
\Theta_1 (z, \tau) }
{ \Theta_1  ( z + (i + \frac{1}{2})\tau , {\bf k} \tau) 
\Theta_1 ( z + (j + \frac{1}{2})\tau , {\bf k} \tau) }, \\
\widetilde{\rm R}_{l,m} (z, \tau) &= 
\frac{ e^{\frac{- \pi \sqrt{-1}}{2}}
q^{H-\frac{Q}{2}} ( e^{- \pi \sqrt{-1}}y)^{Q - 
\frac{{\bf c}-1}{2}}
 \Theta_1 ((i-j)\tau, {\bf k} \tau) 
\Theta_1 (z, \tau) }
{ \Theta_1  ( z + (i - \frac{1}{2})\tau , {\bf k} \tau) 
\Theta_1 ( z + (j - \frac{1}{2})\tau , {\bf k} \tau) } \ .
\elea(Rch)
By (\req(threl)) (\req(chs)), and the relations
$$
1 \leq i+\frac{1}{2} \leq k+1 \ , \ -k \leq j+\frac{1}{2} \leq 0  \ , 
\ \ 0 \leq i-\frac{1}{2} \leq k \ , 
\ \ , \ -(k+1) \leq j-\frac{1}{2} \leq -1 \ , 
$$
one obtains the following relations of Ramon 
characters, which was used in the study of
topological elliptic genus \cite{EOTY,R} 
\par \vspace{0.1in} \noindent 
{\bf Corollary 1.}\footnote{ In the paper 
\cite{R}, the signs in the
formula of
$\widetilde{R}_{\lambda}$ on pp. 373
were inaccurate in order to keep the statement of
Proposition 4 there to be valid. The correct
one should be  
$\widetilde{R}_{\lambda}(z, \tau) =
NS_{\lambda} (z-\frac{\tau}{2} -\frac{1}{2},
\tau) {\rm exp}(\frac{c(K)}{3} \pi
\sqrt{-1}(\frac{
\tau}{4} - z+1/2 ))$. Then 
Lemma 4 in \cite{R} holds for $R_{\lambda}$ of this
present paper, instead of
$\widetilde{R}_{\lambda}$. } 
 $$
\begin{array}{l}
{\rm R}_{l, -m} (z, \tau) = 
\widetilde{\rm R}_{l, m} (-z, \tau) \ , \ \ \ 
\  \ \ \ \ \  \ \ \ \ \  \ \ \ \ \  \ \ \ \ \  \
\ \ \  \ \ \ \ 
\widetilde{\rm R}_{l, -m} (z, \tau) = 
{\rm R}_{l, m} (-z, \tau) , \ \\
{\rm R}_{l, m} (- z, \tau)= e^{{\bf
c}\pi\sqrt{-1}  (\tau -2z+1)}
{\rm R}_{l, -m} (z- \tau -1, \tau)   
 \ , \ \ \
\widetilde{\rm R}_{l, m} (- z, \tau)= e^{{\bf c}
\pi\sqrt{-1} (\tau +2z+1)}
\widetilde{\rm R}_{l, -m} (z + \tau + 1 , \tau) \
,
\\ {\rm R}_{l, -m} ( z, \tau)= e^{{\bf
c}\pi\sqrt{-1} (\tau +2z+1)} {\rm R}_{l, m} (-z-
\tau -1,
\tau) \ , \ \ \
\widetilde{\rm R}_{l, -m} ( z, \tau)= e^{{\bf
c}\pi\sqrt{-1} (\tau -2z+1)}
\widetilde{\rm R}_{l, m} (-z + \tau + 1 , \tau) 
\ , \\
{\rm R}_{l,m} (0, \tau) = \left\{ \begin{array}{ll}
(-1)^{Q_{l,m} + \frac{\bf c}{2}} & \mbox{if $m=-l$, i.e. antichiral , } \\
0 & \mbox{otherwise} , 
\end{array}  \right.  \\
\widetilde{\rm R}_{l,m} (0, \tau)  
= \left\{ \begin{array}{ll}
(-1)^{-Q_{l,m} + 
\frac{{\bf c}}{2}} & \mbox{if $m=l$, i.e. chiral, } \\
0 & \mbox{otherwise} \ .
\end{array} \right.
\end{array}
$$
$\Box$ \par \vspace{.2in} \noindent
For characters of other sectors, by (\req(pzPsi))
(\req(threl)) and  Theorem 1, one has the
following result. 
\par \vspace{0.1in} \noindent 
{\bf Corollary 2. } The following relations hold 
for characters of HWMs for a fixed central element
${\bf c}$:
$$
e^{2 \pi \sqrt{-1} Q_{l,m}} {\rm Ch}^{(a,
b)}_{l,m}(z, \tau) = e^{- 2{\bf c}
\pi \sqrt{-1}a} {\rm Ch}^{(a, b)}_{l,m}(z+ 1,
\tau)
 \ , \ \ \
{\rm Ch}^{(a,b)}_{l, m-2}(z, \tau) = e^{{\bf c} 
\pi \sqrt{-1}(\tau + 2(z+b))} 
{\rm Ch}^{(a, b)}_{l,m}(z+\tau, \tau) \ .
$$
$\Box$ \par \vspace{.2in} \noindent 
For the qualitative study of 
characters,  it is more convenient to extend the
domain of
$(l, m)$ to the  following  larger one: 
$$
{\cal L}_{\bf c} := \{ (l, m) \in \ZZ^2 \ | \ 
0 \leq l \leq {\bf k}-2 , \ l \equiv m \pmod{2} \} \ ( 
\Longleftrightarrow  
\{ (i, j) \in (\frac{1}{2}+\ZZ)^2 \ | \ 
0 < i-j < {\bf k} \} \ )  \ .
$$
For  $(l,m) \in {\cal L}_{\bf c}$,
one defines  
$
H_{l,m}, Q_{l,m},  {\rm Ch}_{l,m}^{(a, b)}(z,
\tau) $ etc.  by  
the same expressions in  (\req(DS)) 
(\req(chij)), (\req(chtheta)), (\req(Rch)).  The
chirality  now becomes,
$$
{\rm chiral } \Longleftrightarrow 
e^{2 \pi \sqrt{-1}(H_{(l,m)}-
\frac{Q_{(l,m)}}{2})} = 1 \ , \ \ \  {\rm
antichiral } \Longleftrightarrow  e^{2 \pi
\sqrt{-1}(H_{(l,m)}+ \frac{Q_{(l,m)}}{2})} = 1
$$
Define the 
automorphism of ${\cal L}_{\bf c} $,
$$
r: {\cal L}_{\bf c}  \longrightarrow {\cal L}_{\bf c}  \ , \ \
(l, m) \mapsto ({\bf k}-2 -l, m+{\bf k}) \ \ (
\Longleftrightarrow  (i, j) \mapsto (j, i) - (0,
{\bf k}) \ ) . 
$$
Denote ${\cal B}_{\bf c} $ be the space of  
$<r>$-orbits ,  ${\cal V}_{\bf c} $ the  
Hilbert space with ${\cal B}_{\bf c} $ as an orthonormal basis, 
$$
{\cal B}_{\bf c}  = {\cal L}_{\bf c} /<r> \ , \ \ \ \ 
{\cal V}_{\bf c} = \oplus_{\lambda \in {\cal B}_{\bf c} } \CZ \lambda \ .
$$
A fundamental region for ${\cal B}_{\bf c} $ 
consists of those 
$(l,m)$ with $|m|\leq l $, i.e., the index set of
HWM. 

$$
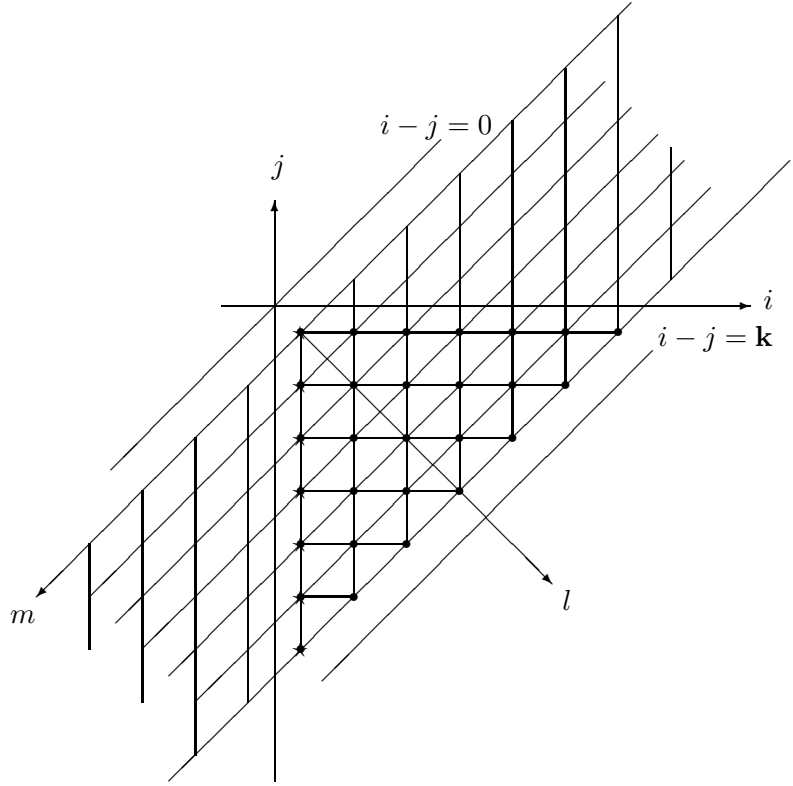
\put(-80,0){\vector(1,0){200}}
\put(125, -1){\shortstack{$i$}}
\put(-60,-180){\vector(0,1){220}}
\put(-61, 50){\shortstack{$j$}}
\put( -50, -10){\vector(1,-1){95}}
\put( 49, -115){\shortstack{$l$}}
\put( 75, 115){\vector(-1,-1){225}}
\put( -160, -120){\shortstack{$m$}}
\put( -20, 65){\shortstack{$i-j =0$}}
\put( 3, 63){\line(-1,-1){125}}
\put( 65, 85){\line(-1,-1){195}}
\put( 75, 75){\line(-1,-1){195}}
\put( 85, 65){\line(-1,-1){195}}
\put( 95, 55){\line(-1,-1){195}}
\put( 105, 45){\line(-1,-1){195}}
\put( 135, 55){\line(-1,-1){235}}
\put( 83, -17){\line(-1,-1){125}}
\put( 85, -15){\shortstack{$i-j ={\bf k}$}}
\put(-53, -13){\shortstack{$\star$}}
\put(-53, -33){\shortstack{$\star$}}
\put(-53, -53){\shortstack{$\star$}}
\put(-53, -73){\shortstack{$\star$}}
\put(-53, -93){\shortstack{$\star$}}
\put(-53, -113){\shortstack{$\star$}}
\put(-53, -133){\shortstack{$\star$}}
\put(-50, -10){\circle*{3}}
\put(-30, -30){\circle*{3}}
\put(-10, -50){\circle*{3}}
\put(10, -70){\circle*{3}}
\put(-50, -30){\circle*{3}}
\put(-30, -10){\circle*{3}}
\put(-50, -50){\circle*{3}}
\put(-10, -10){\circle*{3}}
\put(-50, -70){\circle*{3}}
\put(10, -10){\circle*{3}}
\put(-50, -90){\circle*{3}}
\put(30, -10){\circle*{3}}
\put(-50, -110){\circle*{3}}
\put(50, -10){\circle*{3}}
\put(-50, -130){\circle*{3}}
\put(70, -10){\circle*{3}}
\put(-30, -50){\circle*{3}}
\put(-30, -70){\circle*{3}}
\put(-30, -90){\circle*{3}}
\put(-30, -110){\circle*{3}}
\put(-10, -30){\circle*{3}}
\put(-10, -70){\circle*{3}}
\put(-10, -90){\circle*{3}}
\put(10, -50){\circle*{3}}
\put(10, -30){\circle*{3}}
\put(30, -30){\circle*{3}}
\put(30, -50){\circle*{3}}
\put(50, -30){\circle*{3}}
\put(-130,-90){\line(0,-1){40}}
\put(-110,-70){\line(0,-1){80}}
\put(-90,-50){\line(0,-1){120}}
\put(-70,-30){\line(0,-1){120}}
\put(-50,-10){\line(0,-1){120}}
\put(-30, 10){\line(0,-1){120}}
\put(-10, 30){\line(0,-1){120}}
\put(10, 50){\line(0,-1){120}}
\put(30, 70){\line(0,-1){120}}
\put(50, 90){\line(0,-1){120}}
\put(70, 110){\line(0,-1){120}}
\put(90, 60){\line(0,-1){50}}
\put(-50,-10){\line(1,0){120}}
\put(-50,-30){\line(1,0){100}}
\put(-50,-50){\line(1,0){80}}
\put(-50,-70){\line(1,0){60}}
\put(-50,-90){\line(1,0){40}}
\put(-50,-110){\line(1,0){20}}
$$
Note that 
the characters and chirality thus defined are
invariant under the   automorphism $r$, 
i.e.,  depending only on the class 
 of $(l, m)$ 
in ${\cal B}_{\bf c} $, which will be 
denoted by 
$[l,m]$. We
denote  
$$
{\rm Ch}_{\lambda}^{(a, b)}(z, \tau) := {\rm
Ch}_{l,m}^{(a, b)}(z, \tau) \ \ \  \  {\rm for} \
\
\lambda= [l, m] \in {\cal B}_{\bf c} \ ,
$$ 
similar for ${\rm NS}_{\lambda}(z, \tau), 
{\rm R}_{\lambda}(z, \tau), \widetilde{\rm R}_{\lambda}(z,
\tau)$. We introduce the following automophisms of
${\cal L}_{\bf c} $, 
$$
\begin{array}{ll}
{\sf f}_0: (l, m) \mapsto (l, m-2) \ &( \Longleftrightarrow 
(i, j) \mapsto (i+1, j+1)  \ ) \ , \\
{\sf n}_0: (l, m) \mapsto (l, -m) \ &( \Longleftrightarrow 
(i, j) \mapsto (-j, -i)  \ ) \ .
\end{array} 
$$
By ${\sf f}_0 r = r {\sf f}_0 ,  {\sf n}_0 r = r^{-1} 
{\sf n}_0 $, the maps
${\sf f}_0, {\sf n}_0$ induce the one-one
correspondences of 
${\cal B}_{\bf c} $, hence   
the linear automorphisms ${\sf f}, {\sf n}$ of the
vector space 
${\cal V}_{\bf c}$  with the order ${\bf
k},  2$ respectively, 
\bea(l)
{\sf f} : {\cal V}_{\bf c} \longrightarrow {\cal V}_{\bf c} \ , \ 
\ \ \ [l, m] 
\in {\cal B}_{\bf c}  \ \  \mapsto 
[l, m-2] \in {\cal B}_{\bf c}   \ , \\
{\sf n} : {\cal V}_{\bf c} \longrightarrow {\cal V}_{\bf c} \ , \ 
\ \ \ [l, m] 
\in {\cal B}_{\bf c}  \ \  \mapsto 
[l, -m]\in {\cal B}_{\bf c}   \ .
\elea(flow)
The linear transformations ${\sf f}, {\sf n}$
are called the spectral flow, the charge 
conjugation of
${\cal V}_{\bf c}$ respectively. By Corollary 1 of
Theorem 1, one has the following relations of 
Ramon  characters:
\bea(ll)
{\rm R}_{{\sf n} (\lambda)} (z, \tau) = \widetilde{\rm R}_{
\lambda} (-z, \tau), &
\widetilde{\rm R}_{{\sf n} (\lambda)} (z, \tau) = {\rm R}_{\lambda} (-z, \tau) , \ \\
{\rm R}_{\lambda} (- z, \tau)= e^{{\bf
c}\pi\sqrt{-1} (\tau -2z+1)} {\rm R}_{{\sf n}
(\lambda)} (z- \tau -1, \tau) ,  &
\widetilde{\rm R}_{\lambda} (- z, \tau)= e^{{\bf
c}\pi\sqrt{-1} (\tau +2z+1)}
\widetilde{\rm R}_{{\sf n} (\lambda)} 
(z + \tau + 1 , \tau) \ , \\
{\rm R}_{{\sf n} (\lambda)} ( z, \tau)= e^{{\bf
c}\pi\sqrt{-1} (\tau +2z+1)} {\rm R}_{\lambda}
(-z- \tau -1, \tau) ,  &
\widetilde{\rm R}_{{\sf n} (\lambda)} ( z, \tau)= e^{{\bf
c}\pi\sqrt{-1} (\tau -2z+1)}
\widetilde{\rm R}_{\lambda} (-z + \tau + 1 ,
\tau) \ . 
\elea(inR)
By Corollary 2 of Theorem 1, we have
the following  
relations of  ${\rm Ch}^{(a,b)}_{\lambda}(z,
\tau)$   for $\lambda \in {\cal B}_{\bf c} $, by
which  the elliptic property of characters will
be discussed in the next section,
\bea(l)
 e^{{\bf c} 
\pi \sqrt{-1}(\tau + 2(z+b))} {\rm Ch}^{(a,
b)}_{\lambda}(z+\tau, \tau) = {\rm
Ch}^{(a,b)}_{{\sf f}(\lambda)}(z, \tau) \ , \ \ 
 e^{- 2{\bf c}
\pi \sqrt{-1}a} {\rm Ch}^{(a,b)}_{\lambda}(z+ 1,
\tau) =  e^{2 \pi \sqrt{-1} Q_{\lambda}} {\rm
Ch}^{(a, b)}_{\lambda} (z, \tau) 
 \ ,
\elea(Chll)
in particular, for the Neveu-Schwarz and Ramon 
characters one has 
\bea(ll)
e^{{\bf c} 
\pi \sqrt{-1}(\tau + 2z)} {\rm
NS}_{\lambda}(z+\tau, \tau) = {\rm NS}_{{\sf
f}(\lambda)}(z, \tau) \ , & {\rm NS}_{\lambda}(z+
1, \tau) =  e^{2 \pi \sqrt{-1} Q_{\lambda}} {\rm
NS}_{\lambda} (z, \tau) 
 \ ,  \\
 e^{{\bf c} 
\pi \sqrt{-1}(\tau + 2z+1)} {\rm
R}_{\lambda}(z+\tau, \tau) = {\rm R}_{{\sf
f}(\lambda)}(z, \tau) ,&e^{- {\bf c}
\pi \sqrt{-1}} {\rm R}_{\lambda}(z+ 1, \tau) = 
e^{2 \pi \sqrt{-1} Q_{\lambda}} {\rm R}_{\lambda}
(z, \tau) 
 \ , \\
 e^{{\bf c} 
\pi \sqrt{-1}(\tau + 2z-1)} \widetilde{\rm
R}_{\lambda}(z+\tau, \tau) = \widetilde{\rm
R}_{{\sf f}(\lambda)}(z, \tau) \ ,& e^{{\bf c}
\pi \sqrt{-1}} \widetilde{\rm R}_{\lambda}(z+ 1,
\tau) =  e^{2 \pi \sqrt{-1} Q_{\lambda}}
\widetilde{\rm R}_{\lambda} (z, \tau) 
 \ . 
\elea(NSRll)

\section{Ellipticity and Modularity of Superconformal Algebra Representations}
In this section, we are going to discuss the 
symmetry properties among the unitary  irreducible
HWMs with a fixed central
element 
${\bf c}= \frac{{\bf k}-2}{\bf k}$ for  ${\sf
SCA}$-modules. First we describe some notions which
will facilitate  the later presentation
significantly.  For a positive  integer $N$, a
vector
$v$  in $\CZ^N$ will be
identified with a sequence of complex numbers, 
$(v_m)_{m\in \ZZ}$, with $v_m = v_{m+N}$. The
standard basis of
$\CZ^N$  are denoted by $|n>, n \in \ZZ$ with 
$|n> = |m>$ if $n \equiv m \pmod{N}$.  Denote
${\rm V}_N, {\rm U}_N$ and ${\rm U}_N^{\zeta} \
(\zeta
\in \CZ^*)$ as the 
linear automorphisms of $\CZ^N$ defined by
\begin{eqnarray*}
&{\rm V}_N|n> = \omega^n |n > \ , \ \ \ \ \ \ 
{\rm U}_N|n> = |n+1>
\ , \\
&{\rm U}_N^{\zeta}|n> =
\left\{\begin{array}{ll}
\zeta |n+1> & {\rm if} \  n \equiv 0 \pmod{N},
\\ |n+1> & {\rm otherwise}.
\end{array} \right.
\end{eqnarray*}
Here $\omega$ is a primitive $N$-th root of
unity, which we will specify later. 

For a given $r=(a, b)$, with the identification
of  the basis element 
$\lambda$ of ${\cal V}_{\bf c}$ and 
the character ${\rm
Ch}_{\lambda}^{(a, b)}(z,
\tau)$, the  relation (\req(Chll)) 
gives rise to a $\Lambda({\bf
c})$-representation on 
${\cal V}_{\bf c}$ which describes the elliptic
properties of the HWMs. This $\Lambda({\bf
c})$-representation is
trivial on the 
subgroup
${\bf k}\Lambda$ of  
$\Lambda({\bf c})$ which is generated by  
$({\bf k}, 0)$ and $(0, {\bf k})$, hence it
descends to a representation of the finite
quotient  group,
$\overline{\Lambda({\bf c})}: =
\Lambda({\bf c})/ {\bf k}\Lambda$, 
\bea(l)
 {\cal V}_{\bf c} \times \overline{\Lambda({\bf c})} \longrightarrow {\cal V}_{\bf c} \ .
\elea(cellp)
For easier description of the above
representation, 
we define a pair of integers $N, M$  associated
to the central element ${\bf c}=\frac{{\bf
k}-2}{\bf k}$, 
$$
(N, M) ( = (N({\bf c}), M ({\bf c})) = \left\{ \begin{array}{ll}
({\bf k} , \frac{{\bf k}-1}{2}) & \mbox{for odd } {\bf k} , \\
(\frac{\bf k}{2} , {\bf k}-1)  & \mbox{for even } {\bf k} .
\end{array} \right.  
$$
Denote ${\rm P}=
\overline{(1,0)}, {\rm Q}=\overline{(0,1)}$,
the generators  of 
$\overline{\Lambda({\bf c})}$, and they 
satisfy the Weyl commutation relation:
\bea(l)
{\rm P} {\rm Q} = \omega^{- 1} {\rm Q} {\rm P} \ , 
\ \ \ \ \ 
 \omega: = e^{-2 \pi \sqrt{-1} {\bf c}}  = e^{
\frac{4 \pi \sqrt{-1}}{\bf k}}
\elea(PQ)
Note that 
$\omega$  is 
a primitive $N$-th  root of unity. It is known
that the irreducible unitary  representations
of
$\overline{\Lambda({\bf c})}$ are 
parametrized by elements $(\zeta_1 , \zeta_2)$ in
the (real) 2-torus $\CZ^*_1 \times \CZ^*_1$,
where  
$(\zeta_1 , \zeta_2)$  represents the $N$-dimensional 
$\Lambda ({\bf c})$-module 
$\CZ^N_{(\zeta_1 , \zeta_2)}$ with
${\rm Q} \mapsto \zeta_1^{\frac{1}{N} }{\rm V}_N, 
{\rm P} \mapsto \zeta_2^{\frac{1}{N} }{\rm U}_N$. 
Note that $\CZ^N_{(\zeta_1 , \zeta_2)}$ 
is equivalent to the representation  with $
{\rm Q} \mapsto \zeta_1^{\frac{1}{N} } {\rm V}_N
,  {\rm P} \mapsto {\rm U}_N^{\zeta_2} $.
\par \vspace{.2in} \noindent 
{\bf Lemma 2.} The $\overline{\Lambda({\bf
c})}$-module 
${\cal V}_{\bf c}$ is the following direct sum of 
$M$  irreducible representations, 
$$
{\cal V}_{\bf c} = \left\{ \begin{array}{ll}
M \CZ^N_{(1,1)} & \mbox{for \ odd \ } \ {\bf k}, 
\\
\bigoplus_{\epsilon, \epsilon' = \pm 1} M_{\epsilon, \epsilon'} 
\CZ^N_{(\epsilon, \epsilon')} & \mbox{for \ even \
} {\bf k} , 
\end{array} \right.
$$
where the multiplicities $M_{\epsilon, \epsilon'}$
for even ${\bf k}$ are defined by
$$
\left\{ \begin{array}{lll}
M_{\epsilon, \epsilon'}= \frac{{\bf k}-2}{4} 
& {\rm except} \ 
M_{1,1}= \frac{{\bf k}+2}{4} \ , 
 & \mbox{for }  \frac{{\bf k}}{2} \equiv 1 
\pmod{2} \ , \\
M_{\epsilon, \epsilon'}= 
\frac{{\bf k}}{4} & {\rm except} \
M_{-1,-1}=
\frac{{\bf k}-4}{4} \ ,   & \mbox{for} \ 
 \ \frac{{\bf k}}{2} \equiv 0 \pmod{2}  \ .
\end{array} \right.
$$
\par \vspace{.1in} \noindent
{\it Proof.} The ${\sf f}$-action 
(\req(flow)) on the basis 
${\cal B}_{\bf c} $ of ${\cal
V}_{\bf c}$ gives rise to the $<{\sf
f}>$-orbit decomposition,
$$
{\cal B}_{\bf c}  = 
\bigsqcup_{\nu =0}^{ [ \frac{{\bf k}-2}{2}] }
{\cal B}_{{\bf c}, \nu} \ , \ 
\ {\cal B}_{{\bf c} , \nu} : = \{ [l, m] \in
{\cal B}_{\bf c}  \ | \  l=\nu \} \ .  
$$    
Define ${\cal V}_{{\bf c}, \nu} = \oplus_{\lambda \in {\cal B}_{{\bf c} , \nu}}
\CZ \lambda$. As $\Lambda({\bf c})$-modules, 
${\cal V}_{\bf c}$ is the  direct sum of 
${\cal V}_{{\bf c}, \nu}$s, each of which  is 
${\bf k}$-dimensional except 
dim.${\cal V}_{{\bf c}, \frac{{\bf k}-2}{2}}= \frac{\bf k}{2}$.
For odd 
${\bf k}$,   by (\req(Chll)) the
$\overline{\Lambda ({\bf c})}$-representation 
${\cal V}_{{\bf c}, 
\nu}$ is irreducible, and 
equivalent  to 
$\CZ^N_{(1,1)}$ for all $\nu$. For even ${\bf
k}$,  the   irreducible decomposition of the $
\overline{\Lambda ({\bf c})}$-module  
${\cal V}_{{\bf c}, \nu}$ is given by
$$
{\cal V}_{{\bf c}, \nu } =  \left\{ \begin{array}{ll}
 \CZ^N_{(1,1)} + \CZ^N_{(1,-1)} , & \mbox{if } \ 
0 \leq \nu <  \frac{{\bf k}-2}{2} \ ,  \ \nu 
\equiv 0 \pmod{2} , \\
\CZ^N_{(-1,1)}+ \CZ^N_{(-1,-1)} , & \mbox{if } \ 
0 \leq \nu <  \frac{{\bf k}-2}{2} \ ,
 \ \nu \equiv 1 \pmod{2} \ , \\
\CZ^N_{(1,1)} \ , & \mbox{if } \ 
 \nu =  \frac{{\bf k}-2}{2} \ ,  \ 
\frac{\bf k}{2}  \equiv 1 \pmod{2} , \\
\CZ^N_{(-1,-1)} \ , & \mbox{if } \ 
 \nu =  \frac{{\bf k}-2}{2} \ ,  \ 
\frac{\bf k}{2}  \equiv 0 \pmod{2} .
\end{array} \right.
$$
Then the result follows immediately.
$\Box$ \par \vspace{.2in} \noindent 
For the matrix realization of 
elliptic and modular transformations, we shall
make the identification of vector spaces,  
$\CZ^{M} \otimes \CZ^N = {\cal V}_{\bf c}$, through
a canonical correspondence of the basis elements.
For odd
${\bf k}$,  the identification is made by the
following relations for $0 \leq
\nu <M,  n \in \ZZ$, 
\bea(l)
|\nu> \otimes |n> =  \left\{ \begin{array}{ll}
[\nu, -2n] & \mbox{for even } \ \nu , \\

[\nu, {\bf k}-2n] & \mbox{for odd } \ \nu .
\end{array} \right.
\elea(koe) 
For even ${\bf k}$, with
$\CZ^{M}  = \oplus_{\epsilon, \epsilon'= \pm1 } 
\CZ^{M_{\epsilon, \epsilon'}} $ and the 
standard base of $\CZ^{M_{\epsilon,
\epsilon'}}$ 
denoted     by $|\nu>_{\epsilon, \epsilon'}$
for 
$0 \leq \nu < M_{\epsilon, \epsilon'}$,
the identification of $\CZ^{M} \otimes \CZ^N$ and 
${\cal V}_{\bf c}$ 
is given by 
\bea(lll)
|\nu>_{1,1} \otimes |n> & =  \frac{1}{2} ( 
 [2\nu,-2n] + 
[2\nu, -2n-{\bf k}] ), & \mbox{for \ }
0 \leq \nu  < M_{1,1} , \\ 
|\nu>_{1,-1} \otimes |n> &= \frac{1}{2} (
[2\nu,-2n] - 
[2\nu, -2n-{\bf k}]),  & \mbox{for \ } 
0  \leq \nu < M_{-1,1} ,  \\
|\nu>_{-1, 1} \otimes |n> &= \frac{1}{2}( 
[2\nu+1, 1 -2n] + 
[2\nu+1, 1 -2n-{\bf k}]) , & \mbox{for
\ } 0 \leq \nu  < M_{-1,1} , \\
|\nu>_{-1,-1} \otimes |n>
&=\frac{1}{2}( [2\nu+1,1-2n] - 
[2\nu+1, 1-2n-{\bf k}]), & 
\mbox{for \ }  0 \leq \nu < M_{-1,-1} .
\elea(kee)   
Note that by $\overline{
(\frac{{\bf k}-2}{2}, m)}= \overline{
(\frac{{\bf k}-2}{2}, m+{\bf k})} $, one has the 
following identities,
$$ 
\begin{array}{lll}
|\nu>_{1,1} \otimes |n> &=
[2\nu , -2n] ,  & \mbox{for \ } 
 \nu = M_{1,1}-1  \ \ \frac{\bf k}{2} \equiv
1 \ \pmod{2} , \\
|\nu>_{-1, 1} \otimes |n> &= 
[2\nu+1,1-2n] , & \mbox{for \ } 
 \nu = M_{-1,1}-1  \frac{\bf k}{2} \equiv 0 \
\pmod{2} 
\end{array} 
$$
As an example, the 
identification (\req(kee) ) for ${\bf k}=4$ is
given by
$$
\begin{array}{ll}
 |0>_{1,1} \otimes |0> = \frac{1}{2}
[0,0] +[0, -4], & 
|0>_{1,1} \otimes |1> = \frac{1}{2}(
[0,-2] + [0, -6]) ,  \\ 
|0>_{1,-1} \otimes |0> = \frac{1}{2}(
[0,0] - [0, -4]) , & 
|0>_{1,-1} \otimes |1> =
\frac{1}{2}(
[0,-2] -
[0, -6]) , \\ 
|0>_{-1,1} \otimes |0> =
[1,1],  &  |0>_{-1,1} \otimes |1> = 
[1, -1] .
\end{array}
$$ 
By (\req(Chll)), one obtains the following 
results.
\par \vspace{.2in} \noindent
{\bf Proposition 3.} With  the identification (\req(koe)),  
(\req(kee)) of
${\cal V}_{\bf c} = \CZ^{M} \otimes \CZ^N$,
the 
$\overline{\Lambda({\bf c})}$-representation
(\req(cellp)) on characters  
${\rm Ch}_{\lambda}^{(a,b)}(z, \tau)$ has the
following matrix form:

(i) For odd ${\bf k}, 
{\rm Q} \mapsto {\rm Id} \otimes {\rm V}_N^{-1}  \
, \ \   {\rm P} \mapsto {\rm Id} \otimes {\rm
U}_N$. 

(ii) For even ${\bf k}$, one has the 
$\overline{\Lambda({\bf c})}$-decomposition
$\CZ^M \otimes \CZ^N = \oplus_{\epsilon, \epsilon'=\pm1}
(\CZ^{M_{\epsilon, \epsilon'}} \otimes \CZ^N ) $
with the matrix expressions on $\CZ^{M_{\epsilon,
\epsilon'}} \otimes \CZ^N $ given by   
$$
\begin{array}{lll}
{\rm on} \ \CZ^{M_{1,\epsilon'}} \otimes \CZ^N : &
{\rm Q} \mapsto {\rm Id} \otimes {\rm V}_N^{-1} \
, &  {\rm P} \mapsto {\rm Id} \otimes 
{\rm U}_N^{\epsilon'} \ , \\
{\rm on} \ \CZ^{M_{-1,\epsilon'}}\otimes \CZ^N: &
{\rm Q} \mapsto 
\omega^{\frac{1}{2}}{\rm Id} \otimes {\rm
V}_N^{-1}  \ , &  {\rm P} \mapsto {\rm Id}
\otimes {\rm U}_N^{\epsilon'} ,
\end{array}
$$
where $\epsilon'= \pm 1$. 
$\Box$ \par \vspace{.2in} \noindent
For the discussion of the
modular properties of the characters ${\rm
Ch}_{\lambda}^{(a, b)}(z,
\tau)$, 
the modular group is ${\Gamma}_q$ with $
q= (\frac{1}{2}-a,
\frac{1}{2}-b)$. Its description 
 is given by  (\req(modat)) with $d= {\bf
c}, w=0$
  while restricting on 
${\Gamma}_q$.  Together with the elliptic 
action (\req(cellp)),  the relations among
characters 
${\rm Ch}_{\lambda}^{(a, b)}(z, \tau)$s  give
rise to a finite dimensional representation of
a group $\overline{\Lambda({\bf c})}
\stackrel{r}{*} \Gamma_q$  through
(\req(HeisJ)) for  
$r=(a, b), d = {\bf c},
w=0$.  Here  the group
$\overline{\Lambda({\bf c})} \stackrel{r}{*} 
\Gamma_q$, is defined by the 
following exact sequence using the 
$\Gamma_q$-conjugation invariance property of 
the subgroup ${\bf k}\Lambda$ of
$\Lambda({\bf c})$, 
$$
0 \longrightarrow {\bf k}\Lambda 
\stackrel{r}{*} \Gamma_q
\longrightarrow  \Lambda({\bf c}) \stackrel{r}{*}
\Gamma_q
\longrightarrow 
\overline{\Lambda({\bf c})} \stackrel{r}{*}
\Gamma_q 
\longrightarrow  0 \ .
$$  
In particular,  $\Gamma_q$ is the full modular
group 
$\Gamma$ for the Ramon characters ${\rm
R}_{\lambda}$ or 
$\widetilde{\rm R}_{\lambda}$;   
$\Gamma_{(\frac{1}{2}, \frac{1}{2})}$ 
for  the Neveu-Schwarz characters ${\rm
NS}_{\lambda}$. In this paper, we shall only
consider  these two special cases, partly due to 
the easy description of  generators of these
modular groups , but also its sufficiency for the
future study on topological applications of
manifolds.  We denote
$$
\Gamma_{(\frac{1}{2}, \frac{1}{2})}^{\rm W}({\bf
c}) : = 
 \overline{\Lambda({\bf c})} * 
\Gamma_{(\frac{1}{2}, \frac{1}{2})} 
\ \ , \ \ \
\Gamma^{\rm W}({\bf c}) : =
\overline{\Lambda({\bf c})}
\stackrel{(\frac{1}{2}, \frac{1}{2})} {*}
\Gamma \ \ .
$$
We now describe the T, S-transforms on 
Ramon and Neveu-Schwarz characters.
\par \vspace{.2in} \noindent
{\bf Proposition 4.} The following relations
hold:
$$
\begin{array}{ll}
{\rm NS}_{\lambda} ( z, \tau + 2) = e^{4 \pi
\sqrt{-1}(H_{\lambda}- 
\frac{\bf c}{8})} 
{\rm NS}_{\lambda} ( z, \tau ) \ ,& 
e^{\frac{-{\bf c} \pi \sqrt{-1} z^2}{\tau}} {\rm
NS}_{\lambda} (\frac{z}{-\tau} , \frac{1}{-\tau})
= 
\sum_{\lambda' \in {\cal B}_{\bf c} } {\rm S}^{\lambda'}_{\lambda} {\rm NS}_{\lambda'} 
(z, \tau) \ , \\
{\rm R}_{\lambda} ( z, \tau + 1) = e^{2 \pi
\sqrt{-1}(H_{\lambda}+ 
\frac{Q_{\lambda}}{2})} 
{\rm R}_{\lambda} ( z, \tau ) \ , & 
e^{ \frac{-{\bf c} \pi \sqrt{-1}z^2}{\tau}}
{\rm R}_{\lambda}
(\frac{z}{-\tau}, \frac{1}{-\tau})  = 
e^{2 \pi \sqrt{-1}(Q_{\lambda}+ \frac{\bf c}{4}
)} 
\sum_{\lambda' \in {\cal B}_{\bf c} } {\rm S}^{\lambda'}_{\lambda} 
{\rm R}_{\lambda'} (z, \tau)  \ , \\ 
\widetilde{\rm R}_{\lambda} ( z, \tau + 1) = e^{2 \pi
\sqrt{-1} (H_{\lambda}- \frac{Q_{\lambda}}{2})} 
\widetilde{\rm R}_{\lambda} ( z, \tau ) \ , &
e^{ \frac{-{\bf c} \pi \sqrt{-1}z^2}{\tau}}
\widetilde{\rm R}_{\lambda}
(\frac{z}{-\tau}, \frac{1}{-\tau})  = 
e^{ 2 \pi \sqrt{-1}(-Q_{\lambda}+ \frac{\bf c}{4}
)} 
\sum_{\lambda' \in {\cal B}_{\bf c} } {\rm S}^{\lambda'}_{\lambda} 
\widetilde{\rm R}_{\lambda'} (z, \tau)  \ ,
\end{array}
$$
where the coefficients 
${\rm S}^{\lambda}_{\lambda'}$, $\lambda = [l, m],
\lambda' = [l', m']$, are defined by
$$
{\rm S}^{\lambda}_{\lambda'} := 
\frac{2 }
{\bf k} \sin \frac{\pi (l+1)(l'+1)}{\bf k} 
e^{\frac{-\pi \sqrt{-1}mm'}{\bf k}}
= \frac{2 }{\bf k} \sin \frac{\pi (i-j)(i'-j')}{\bf k} 
e^{\frac{- \pi \sqrt{-1}(i+j)(i'+j')}{\bf k} } \
 .
$$
\par \vspace{.1in} \noindent 
{\it Proof.} The T-relation of ${\rm R}_{\lambda}, 
\widetilde{\rm R}_{\lambda}$ follows from 
(\req(Rch)) (\req(Psi1m)). By (\req(Ch))  for $(a,
b)=(\frac{1}{2}, 
\frac{1}{2})$ and the T-relation of Ramon 
characters, one obtains the 
${\rm T}^2$-relation of ${\rm NS}_{\lambda}$. 
The S-transform of ${\rm NS}_\lambda$ was known
in literature ,( e.g., formula (28a)
of \cite{RY}), which  we adopt 
here. Then, by
(\req(Ch)), we have the  relation,
$$
e^{ \frac{-{\bf c} \pi \sqrt{-1}z^2}{\tau}}
{\rm Ch}_{\lambda}^{(a,b)} 
(\frac{z}{-\tau}, \frac{1}{-\tau})  = 
e^{2ab{\bf c} \pi \sqrt{-1}}
\sum_{\lambda' \in {\cal B}_{\bf c} } {\rm S}^{\lambda'}_{\lambda} 
{\rm Ch}^{(-b,a)}_{\lambda'} (z, \tau) \ .
$$
Setting  $(a, b) =(\frac{1}{2}, \frac{-1}{2})$ or 
$(\frac{-1}{2}, \frac{1}{2})$ in the above
equality,  
 one  obtains the expression of the S-transform
of  Ramon characters by
(\req(chs)) (\req(NSRll)). 
$\Box$ \par \vspace{.2in} \noindent
With the 
identification of the basis element $\lambda$ of
${\cal V}_{\bf c}$ with the character
${\rm NS}_{\lambda}$ or 
${\rm R}_{\lambda}, \widetilde{\rm
R}_{\lambda}$, by (\req(cellp)), by using
Proposition 4 and  the remark of Proposition 2,
there exists a canonical Jacobi-group
representation on
${\cal V}_{\bf c}$ for the  Neveu-Schwarz or Ramon
sector,
$$
\begin{array}{l}
\Pi_{\rm NS} : {\cal V}_{\bf c} \times 
\Gamma_{(\frac{1}{2}, \frac{1}{2})}^{\rm W}
({\bf c}) 
\longrightarrow {\cal V}_{\bf c} \ , \ \ \ \ \ \ \
\Pi_{\rm R}, \Pi_{\widetilde{\rm R}} :  {\cal V}_{\bf c} \times 
\Gamma^{\rm W}({\bf c})
\longrightarrow {\cal V}_{\bf c} \ .
\end{array}
$$
With ${\cal V}_{\bf c} =
\CZ^{M}
\otimes
\CZ^N$  through (\req(koe))  (\req(kee)), the
matrix  presentation of the above
actions on the part $\overline{\Lambda({\bf c})}$
is described by Proposition 3. As any modular
transformation sends a
$\overline{\Lambda({\bf c})}$-irreducible
subspace onto another one,  we shall be
able to obtain a matrix form
of a modular transformation as a tensor product
of one from the
$\CZ^M$-factor, $ (
\alpha_{\nu
\nu'})_{\nu,
\nu'}$, and another one from the 
$\CZ^N$-factor,  $ ( \beta_{n
n'})_{n, n'}$. By Proposition 4, 
 ${\rm T}$-transformation is a diagonal
matrix; while the ${\rm S}$-transform of 
a vector will be a linear combination of  
all the basis elements.  The explicit forms are
given in the following theorem,  where the
matrix indices $\nu, \nu', n, n'$  are
non-negative integers less than the dimensions of 
vector spaces involved in the expression,.
\par \vspace{.2in} \noindent
{\bf Theorem 2.} For  $\Pi_{\rm NS}$,$\Pi_{\rm
R}, 
\Pi_{\widetilde{\rm R}}$-representations,     
the  matrix forms of the modular
transformations are as follows:

(i) For odd ${\bf k}$, with 
${\cal V}_{\bf c} = \CZ^{M} \otimes \CZ^N$ by
(\req(koe)), we have
$$
\begin{array}{lll}
\Pi_{\rm NS} : &  
{\rm T}^2 &\mapsto
-\sqrt{-1} ( (-1)^{\frac{(-1)^\nu-1}{2}}
e^{ \frac{(\nu +1)^2 \pi \sqrt{-1}}{\bf k}}
\delta_{\nu,\nu'} )
\otimes ( \omega^{-n^2} 
\delta_{n,n'})   \ , \\
& 
{\rm S} &\mapsto  \frac{2 }{\bf k} 
 ( \sin \frac{(\nu+1)(\nu'+1)\pi }{\bf k} ) 
\otimes (
    \omega^{-nn'} )  \ , \\
\Pi_{\rm R} \  : & 
{\rm T} &\mapsto 
( (-1)^{\frac{(-1)^\nu-1}{4}}
 e^{  \frac{(\nu+1)^2 \pi
\sqrt{-1}}{2\bf k}} \delta_{\nu,\nu'}) 
\otimes (
\omega^{\frac{-(2n+1)^2}{ 8 }} \delta_{n,n'}) \ ,
\\
&{\rm S} & \mapsto  \frac{2
\sqrt{-1}}{\bf k}
 ( \sin \frac{(\nu+1)(\nu'+1)\pi }{\bf k} ) 
\otimes (
    \omega^{-n(n'+1)-\frac{1}{4}} ) \ ,  \\
\Pi_{\widetilde{R}} \ : & 
{\rm T} &\mapsto 
( (-1)^{\frac{(-1)^\nu-1}{4}} e^{ 
\frac{(\nu+1)^2 \pi \sqrt{-1}}{2\bf k}}
\delta_{\nu,\nu'}) 
\otimes (
\omega^{\frac{-(2n-1)^2}{ 8 }}\delta_{n,n'}) \ ,
\\

& {\rm S} &\mapsto  \frac{2
\sqrt{-1}}{\bf k}
 ( \sin \frac{(\nu+1)(\nu'+1)\pi }{\bf k} ) 
\otimes ( \omega^{-n(n'-1)-\frac{1}{4}} ) \ ,
\end{array}
$$
where $(-1 )^* := e^{*({\bf k}-2)\pi \sqrt{-1}} $.

(ii) For even ${\bf k}$, with  
${\cal V}_{\bf c} = \oplus_{\epsilon,
\epsilon'=1,2} \CZ^{M_{\epsilon,
\epsilon'}}
\otimes
\CZ^N$ by (\req(kee)), the linear transformations,
${\rm T}, {\rm S}$, are composed of 
 the following ones on the
Neveu-Schwarz, or Ramon sector: 
$$
\begin{array}{ll}
\Pi_{\rm NS}  \ , & {\rm T}^2:
\CZ^{M_{\epsilon,\epsilon'}}
\otimes \CZ^N \longrightarrow 
\CZ^{M_{\epsilon,\epsilon'}} \otimes \CZ^N
\ , \ \ \\
\Pi_{\rm R} , 
\Pi_{\widetilde{\rm R}} \ ,&
{\rm T}: \CZ^{M_{\epsilon,\epsilon'}}
\otimes \CZ^N \longrightarrow 
\CZ^{M_{\epsilon, \epsilon''}} \otimes \CZ^N \ ,
 \\
\Pi_{\rm NS}, \Pi_{\rm R}, 
\Pi_{\widetilde{\rm R}} \ , &
{\rm S}: \CZ^{M_{\epsilon,\epsilon'}}
\otimes \CZ^N \longrightarrow 
\CZ^{M_{\epsilon', \epsilon}} \otimes \CZ^N \ ,
\end{array}
$$
where $\epsilon, \epsilon'= \pm 1$, and 
$
\ \epsilon'': =  (-1)^{\frac{{\bf
k}-2}{2}}\epsilon
\epsilon'$ . 
The matrix representation of the above linear 
transformations is given by
$$
\begin{array}{lll}
\Pi_{\rm NS} , &
\CZ^{M_{\epsilon,\epsilon'}}
\otimes \CZ^N \stackrel{{\rm T}^2
}{\longrightarrow} 
\CZ^{M_{\epsilon,\epsilon'}} \otimes \CZ^N \ , 
& \mapsto   -\sqrt{-1} ( 
e^{\frac{(4\nu+3 -\epsilon)^2\pi \sqrt{-1}}{4\bf
k }}
\delta_{\nu,\nu'}) \otimes 
(\omega^{\frac{
-(4n-1+\epsilon)^2}{16}}
\delta_{n,n'})
\ ,
\\
\Pi_{\rm R} , &
\CZ^{M_{\epsilon,\epsilon'}}
\otimes \CZ^N \stackrel{\rm T
}{\longrightarrow} 
\CZ^{M_{\epsilon,\epsilon''}} \otimes \CZ^N \ , 
& \mapsto 
( e^{\frac{(4\nu+3 -\epsilon)^2\pi
\sqrt{-1}}{8\bf k }}
\delta_{\nu,\nu'}) \otimes (
\omega^{\frac{-(4n +1 + \epsilon)^2}{32} }
\delta_{n,n'}) \  , \\
\Pi_{\widetilde{\rm R}} , &
\CZ^{M_{\epsilon,\epsilon'}}
\otimes \CZ^N \stackrel{\rm T}{\longrightarrow} 
\CZ^{M_{\epsilon,\epsilon''}} \otimes \CZ^N \ , 
& \mapsto 
( e^{\frac{(4\nu+3 -\epsilon)^2\pi
\sqrt{-1}}{8\bf k }}
\delta_{\nu,\nu'}) \otimes 
(\omega^{\frac{-(4n -3+\epsilon)^2}{32} } 
\delta_{n,n'}) \ ,
\end{array}
$$
and 
$$
\begin{array}{ll}
\CZ^{M_{1.1}} \otimes \CZ^N
\stackrel{{\rm S}}{\longrightarrow}
\CZ^{M_{1,1}} \otimes \CZ^N  ,   & \mapsto 
\left\{ \begin{array}{ll}
\frac{4}{\bf k} (  w_{1,1}(\nu')  \sin
\frac{\pi ( 2\nu+1)(2\nu'+1)}{\bf k}  ) \otimes (
\omega^{nn'})  &{\rm for} \ \Pi_{\rm NS} \ , \\ 
\frac{4 \sqrt{-1}  }
{\bf k}  ( w_{1,1}(\nu') \sin \frac{\pi
( 2\nu+1)(2\nu'+1)}{\bf k}) 
\otimes  (
 \omega^{n(n'-1)- \frac{1}{4}}) 
&{\rm for} \ \Pi_{\rm R} \ , \\
\frac{4\sqrt{-1}  }
{\bf k}  ( w_{1,1}(\nu') \sin \frac{\pi
( 2\nu+1)(2\nu'+1)}{\bf k} ) \otimes  
( \omega^{n(n'+1)-\frac{1}{4}})
 &{\rm for} \ \Pi_{\widetilde{\rm R}} \ ,
\end{array} \right.
\\
 \CZ^{M_{1.-1}} \otimes \CZ^N
\stackrel{{\rm S}}{\longrightarrow} \CZ^{M_{-1,1}}
\otimes 
\CZ^N ,   & \mapsto
\left\{ \begin{array}{ll}
\frac{4}{\bf k} (  w_{-1,1}(\nu') \sin \frac{\pi (
2\nu+1)(2\nu'+2)}{\bf k}   ) \otimes (  
 \omega^{n(n'-\frac{1}{2})}  )  &{\rm for} \
\Pi_{\rm NS} \ , \\ 
\frac{4\sqrt{-1} }{\bf k}  
( w_{-1,1}(\nu') \sin
\frac{\pi ( 2\nu+1)(2\nu'+2)}{\bf k} )
\otimes (
 \omega^{n(n'-\frac{3}{2})-\frac{1}{4}} )
&{\rm for} \ \Pi_{\rm R} \ , \\ 
\frac{4\sqrt{-1} }{\bf k}  
( w_{-1,1}(\nu') \sin
\frac{\pi ( 2\nu+1)(2\nu'+2)}{\bf k}) \otimes
( \omega^{n(n'+\frac{1}{2})-\frac{1}{4}} )
&{\rm for} \ \Pi_{\widetilde{\rm R}} \ ,
\end{array} \right.
\\
 \CZ^{M_{-1.1}} \otimes \CZ^N 
\stackrel{{\rm S}}{\longrightarrow} \CZ^{M_{1,
-1}} \otimes 
\CZ^N ,   & \mapsto 
\left\{ \begin{array}{ll}
\frac{4 }{\bf k}(
\sin \frac{\pi (2\nu +2)(2\nu'+1)}{\bf k} )
\otimes (  \omega^{(n-\frac{1}{2})n'}  )  \ 
&{\rm for} \
\Pi_{\rm NS} \ , \\  
\frac{4 \sqrt{-1}  }
{\bf k} (
\sin \frac{\pi (2\nu +2)(2\nu'+1)}{\bf k})
\otimes
(\omega^{(n-\frac{1}{2})n'-n+\frac{1}{4}})
&{\rm for} \ \Pi_{\rm R} \ , \\ 
\frac{4\sqrt{-1}  }
{\bf k} ( 
\sin \frac{\pi (2\nu +2)(2\nu'+1)}{\bf k} )
\otimes 
(\omega^{(n-\frac{1}{2})n'+n-\frac{3}{4}})
&{\rm for} \ \Pi_{\widetilde{\rm R}} \ ,
\end{array} \right.
\\
 \CZ^{M_{-1.-1}} \otimes \CZ^N 
\stackrel{{\rm S}}{\longrightarrow} \CZ^{M_{-1,
-1}} \otimes 
\CZ^N ,   
& \mapsto
\left\{ \begin{array}{ll}
\frac{4 }{\bf k} ( 
\sin \frac{\pi (2\nu+2)(2\nu'+2)}{\bf k}  )
\otimes ( 
\omega^{(n-\frac{1}{2})(n'-\frac{1}{2})}  ) \
&{\rm for} \
\Pi_{\rm NS} \ , \\ 
\frac{4\sqrt{-1} }
{\bf k}  
(\sin \frac{\pi (2\nu+2)(2\nu'+2)}{\bf k})
\otimes
(\omega^{(n-\frac{1}{2})(n'-\frac{1}{2})-n+
\frac{1}{4}} )
&{\rm for} \ \Pi_{\rm R} \ , \\ 
\frac{4\sqrt{-1} }
{\bf k} 
(\sin \frac{\pi (2\nu+2)(2\nu'+2)}{\bf k})
\otimes 
(\omega^{(n-\frac{1}{2})(n'-\frac{1}{2})+n
-\frac{3}{4}}) \
&{\rm for} \ \Pi_{\widetilde{\rm R}} \ .
\end{array} \right. 
\end{array}
$$
Here the functions, $w_{1,1}(\nu' ), w_{-1,
1}(\nu')$, are defined by
$$
\begin{array}{l}
w_{1,1}(\nu') = \left\{ \begin{array}{ll}
\frac{1}{2} &{\rm if} \ \nu'= M_{1,1}-1 , 
\frac{\bf k}{2}\equiv 1 \pmod{2}, \\
1 &{\rm otherwise \ with \ } 0 \leq \nu' < M_{1,1}
\ ,
\end{array}
\right. \\
w_{-1,1}(\nu') = \left\{ \begin{array}{ll}
\frac{1}{2} &{\rm if} \ \nu'= M_{-1,1}-1, 
\frac{\bf k}{2} \equiv 0 \pmod{2}, \\
1 &{\rm otherwise \ with \ } 0 \leq \nu' <
M_{-1,1}
\ .
\end{array}
\right.
\end{array}
$$
\par \vspace{.1in} \noindent
{\it Proof.} (i) For odd ${\bf k}$, we have 
${\bf k}=N=2M+1$. 
By Proposition 4 and (\req(DS)), one has the 
following expressions of the ${\rm
T}$-transformation on the Neveu-Schwarz or Ramon
sector,
$$
\begin{array}{lll}
\Pi_{\rm NS} \ : & {\rm T}^2 ( |\nu> \otimes |n> )
&
= -\sqrt{-1} (-1)^{\frac{(-1)^\nu-1}{2}} e^{
\frac{(\nu +1)^2 \pi \sqrt{-1}}{\bf k}}
\omega^{-n^2}  |\nu> 
\otimes   |n>  \ , \\ 
&
{\rm S}(|\nu> \otimes |n> )  &= \frac{2 } {\bf k}
\sum_{\nu' = 0}^{M-1} \sum_{ n' =0}^{N-1 }
 \sin \frac{(\nu+1)(\nu'+1)\pi }{\bf k}
    \omega^{-nn'} |{\nu'}> 
\otimes   |n'>  \ , \\
\Pi_{\rm R} \  : & 
{\rm T}(
|\nu> \otimes |n> )
&=(-1)^{\frac{(-1)^\nu-1}{4}} e^{ 
\frac{(\nu+1)^2 \pi \sqrt{-1}}{2\bf k}} 
\omega^{\frac{-(2n+1)^2}{ 8 }}
|\nu> \otimes   |n> \ , \\
&{\rm S} ( |\nu> \otimes |n> )  
&= \frac{2 \sqrt{-1}}
{\bf k} 
\sum_{\nu' = 0}^{M-1} \sum_{ n' =0}^{N-1 }
 \sin \frac{(\nu+1)(\nu'+1)\pi }{\bf k}
    \omega^{-n(n'+1)-\frac{1}{4}} |{\nu'}>
\otimes   |n'> \ , \\
\Pi_{\widetilde{R}} \ :  &
{\rm T}(
|\nu> \otimes |n> ) &
=(-1)^{\frac{(-1)^\nu-1}{4}}
e^{  \frac{(\nu+1)^2 \pi \sqrt{-1}}{2\bf k}} 
\omega^{\frac{-(2n-1)^2}{ 8 }}|\nu> \otimes  
 |n> \ , \\
&{\rm S} ( |\nu> \otimes |n> )  &=
 \frac{2 \sqrt{-1}} {\bf k} 
\sum_{\nu' = 0}^{M-1} \sum_{ n' =0}^{N-1 }
 \sin \frac{(\nu+1)(\nu'+1)\pi }{\bf k}
    \omega^{-n(n'-1)-\frac{1}{4}} |{\nu'}>
\otimes   |n'>  \ .
\end{array}
$$
Then  
the results follow immediately.  

(ii) When
${\bf k}$ is even , we have
${\bf k}= 2N= M+1$, and $\omega^{\frac{\bf k}{
2}}=1$. One has the following
identities for $ \lambda=[l, m] , \lambda'= [l,
m-{\bf k}]$, 
$$
e^{4 \pi \sqrt{-1}(H_{\lambda'})} = e^{4 \pi
\sqrt{-1}(H_\lambda)} \ , \ \ \ e^{2 \pi
\sqrt{-1}(H_{\lambda'}\pm \frac{Q_{\lambda'}}{2})
} =  (-1)^{\frac{{\bf k}-2}{2}+m} e^{2 \pi
\sqrt{-1}(H_{\lambda}\pm \frac{Q_{\lambda}}{2}) }
\ , \ \
\  e^{2 \pi \sqrt{-1}Q_{\lambda'} } = e^{2 \pi
\sqrt{-1}Q_\lambda } \  .
$$
By the first two equalities in the above and 
Proposition 4, one obtains the statement on 
${\rm T}^2$ or ${\rm T}$ on
the $\CZ^{M_{\epsilon \epsilon'}}\otimes
\CZ^N$ for the Neveu-Schwarz, Ramon sector
respectively. Furthermore, by 
$$
\begin{array}{l}
e^{4 \pi \sqrt{-1}(H_{[2\nu, -2n]} - 
\frac{\bf c}{8})} = -\sqrt{-1}
e^{\frac{(2\nu+1)^2\pi
\sqrt{-1}}{\bf k }}
\omega^{-n^2} \ ,  \ \ 
e^{4 \pi \sqrt{-1}(H_{[2\nu+1, 1-2n]} - 
\frac{\bf c}{8})}  = -\sqrt{-1}
e^{ \frac{(2\nu+2)^2
\pi \sqrt{-1} }{ \bf k } }
\omega^{- (n - \frac{1}{2})^2}, \\

e^{2 \pi \sqrt{-1}(H_{[2\nu, -2n]}\pm
\frac{Q_{[2\nu, -2n]}}{2}) } = 
e^{ \frac{(2\nu+1)^2\pi \sqrt{-1}}{2\bf k}}
\omega^{\frac{-(n\pm1/2)^2}{2} } \ ,
 \\
e^{2 \pi \sqrt{-1}(H_{[2\nu+1, 1-2n]} \pm
\frac{Q_{[2\nu+1, 1-2n]}}{2}) } =
e^{ \frac{(2\nu+2)^2 \pi \sqrt{-1}
}{2 \bf k} } \omega^{\frac{ -(n-\frac{1}{2}\pm
\frac{1}{2})^2 }{2 } } \ ,
\end{array}
$$
the expressions of ${\rm
T}$-transformation on the basis elements of 
$\CZ^{M_{\epsilon \epsilon'}}\otimes
\CZ^N$ are as follows, 
$$
\begin{array}{lll}
\Pi_{\rm NS} \ : &{\rm T}^2 (|\nu>_{1,\epsilon'}
\otimes |n>) =  -\sqrt{-1}
e^{\frac{(2 \nu+1)^2\pi \sqrt{-1}}{\bf k }}
\omega^{-n^2} |\nu>_{1,\epsilon'} \otimes |n> \ ,
\\ 
&{\rm T}^2 (|\nu>_{-1,\epsilon'} \otimes |n>) = 
-\sqrt{-1} 
e^{ \frac{(2\nu+2)^2 \pi \sqrt{-1} }{ \bf k } }
\omega^{- (n - \frac{1}{2})^2}
|\nu>_{-1,\epsilon'} \otimes |n> \ , \\
\Pi_{\rm R} \ : &{\rm T} (|\nu>_{1,\epsilon'}
\otimes |n>) =
e^{ \frac{(2\nu+1)^2\pi \sqrt{-1}}{2\bf k}}
\omega^{\frac{-(n +1/2)^2}{2} } 
|\nu>_{1,\epsilon''}
\otimes |n> \ ,  \\
&{\rm T} (|\nu>_{-1,\epsilon'} \otimes |n>) 
=
e^{ \frac{(2\nu+2)^2 \pi \sqrt{-1}
}{2 \bf k} } \omega^{\frac{ -n^2}{2
} } |\nu>_{-1,\epsilon''}
\otimes |n> \ , \\
\Pi_{\widetilde{\rm R}} \ : &{\rm T}
(|\nu>_{1,\epsilon'}
\otimes |n>) =
e^{ \frac{(2\nu+1)^2\pi \sqrt{-1}}{2\bf k}}
\omega^{\frac{-(n -1/2)^2}{2} } 
|\nu>_{1,\epsilon''}
\otimes |n> \ ,  \\
 &{\rm T} (|\nu>_{-1,\epsilon'} \otimes |n>) 
=
e^{ \frac{(2\nu+2)^2 \pi \sqrt{-1}
}{2 \bf k} } \omega^{\frac{ -(n-1)^2 }{2 } }
|\nu>_{-1,\epsilon''}
\otimes |n> \ .
\end{array}
$$
Hence we obtain the matrix expression of ${\rm
T}^2$ for $\Pi_{\rm NS}$; and ${\rm
T}$ for $\Pi_{\rm R},\Pi_{\widetilde{\rm R}}
$-action.  
For the ${\rm S}$-transformation, we
first consider the
Neveu-Schwarz sector.  By the definition of
${\rm S}_{\lambda}^{\lambda'}$, the following
equalities hold: for even $\nu$,  
$$
\begin{array}{ll}
 {\rm S}([\nu,-2n])  &= \frac{2 }
{\bf k}
\sum_{{\rm even } \ \nu' = 0}^{\frac{{\bf
k}-2}{2}-1}
\sin \frac{\pi (
\nu+1)(\nu'+1)}{\bf k} 
\sum_{ n' =0}^{\frac{\bf k}{2}-1 }  \omega^{nn'}
([\nu',  -2n'] 
+ 
[\nu',  -2n'-{\bf k}])\\
&+ \frac{2 }
{\bf k}
\sum_{{\rm odd} \ \nu' = 0}^{\frac{{\bf
k}-2}{2}-1}
\sin \frac{\pi (
\nu+1)(\nu'+1)}{\bf k} 
\sum_{ n' =0}^{\frac{\bf k}{2}-1 }  \omega^{n(n'-\frac{1}{2})}
([\nu',  1-2n']+[\nu',  1-2n'-{\bf
k}] ) \\ &+ 
 \frac{2 }
{\bf k} \sin \frac{\pi (\nu +1)}{2} 
\sum_{ n' =0}^{\frac{\bf k}{2}-1 } \omega^{n(
n'- \frac{\theta({\bf k})}{2})}
[\frac{{\bf k}-2}{2},  \theta({\bf
k})-2n'] \ , \\
 {\rm S}([\nu,-2n-{\bf k}])  &=
\frac{2 } {\bf k} 
\sum_{{\rm even} \ \nu' = 0}^{\frac{{\bf
k}-2}{2}-1}
\sin \frac{\pi (
\nu+1)(\nu'+1)}{\bf k} 
\sum_{ n' =0}^{\frac{\bf k}{2}-1 }  \omega^{nn'}
([\nu',  -2n'] 
+ 
[\nu',  -2n'-{\bf k}] )\\
&- \frac{2 }
{\bf k}
\sum_{{\rm odd} \ \nu' = 0}^{\frac{{\bf
k}-2}{2}-1}
\sin \frac{\pi (
\nu+1)(\nu'+1)}{\bf k} 
\sum_{ n' =0}^{\frac{\bf k}{2}-1 } \omega^{n(n'-\frac{1}{2})}
([\nu',  1-2n']+[\nu',  1-2n'-{\bf
k}] ) \\ &+ \frac{2 }
{\bf k}
 \sin \frac{\pi (\nu +1)}{2} 
\sum_{ n' =0}^{\frac{\bf k}{2}-1 } 
\omega^{n (
n'- \frac{\theta({\bf k})}{2})}
[\frac{{\bf k}-2}{2},  \theta({\bf
k})-2n'] \ ,
\end{array}
$$
where $\theta({\bf k}):=0, 1$ according to
$\frac{\bf k}{2}$  odd or even respectively ; 
for odd $\nu$,
$$
\begin{array}{ll} 
{\rm S}([\nu,1-2n])  &= \frac{2 }
{\bf k}
\sum_{ {\rm even} \  \nu'=0 }^{\frac{{\bf
k}-2}{2}-1} 
\sin \frac{\pi (\nu +1)(\nu'+1)}{\bf k} 
\sum_{ n' =0}^{\frac{{\bf k}}{2}-1 }  \omega^{(n-\frac{1}{2})n'}
( [\nu',  -2n'] -
[\nu',  -2n'-{\bf k}] ) \\
&+ \frac{2 }
{\bf k}
\sum_{{\rm odd } \ \nu' = 0}^{\frac{{\bf
k}-2}{2}-1} 
\sin \frac{\pi (\nu+1)(\nu'+1)}{\bf k} 
\sum_{ n' =0}^{\frac{\bf k}{2}-1 } \omega^{(n-\frac{1}{2})(n'-\frac{1}{2})} (
[\nu',  1-2n']-
[\nu',  1-2n'-{\bf k}] ) \ , \\
 
{\rm S}([\nu,1-2n-{\bf k}])  &= \frac{2 }
{\bf k} 
\sum_{{\rm even} \ \nu' = 0}^{\frac{{\bf
k}-2}{2}-1} 
\sin \frac{\pi (\nu +1)(\nu'+1)}{\bf k} 
\sum_{ n' =0}^{\frac{{\bf k}}{2}-1 } \omega^{(n-\frac{1}{2})n'}
( [\nu',  -2n'] -
[\nu',  -2n'-{\bf k}] ) \\
&- \frac{2 }
{\bf k}
\sum_{{\rm odd} \ \nu' = 0}^{\frac{{\bf
k}-2}{2}-1} 
\sin \frac{\pi (\nu+1)(\nu'+1)}{\bf k} 
\sum_{ n' =0}^{\frac{\bf k}{2}-1 } 
 \omega^{(n-\frac{1}{2})(n'-\frac{1}{2})} (
[\nu',  1-2n']-
[\nu',  1-2n'-{\bf k}] ) \ .
\end{array}
$$
From which one can
derive the following relations,
$$
\begin{array}{ll}
 {\rm S}( |\nu>_{1,1} \otimes |n> )  
 &= 
\frac{4 }
{\bf k}  \sum_{ \nu' = 0}^{M_{1.1}-1} 
\sum_{ n' =0}^{N-1 } w_{1,1}(\nu') \sin \frac{\pi
( 2\nu+1)(2\nu'+1)}{\bf k} 
 \omega^{nn'}|\nu'>_{1,1} 
\otimes   |n'> 
\ , \\ {\rm S}( |\nu>_{1, -1} \otimes |n> )  
&=
\frac{4}{\bf k}  \sum_{ \nu' = 0}^{M_{-1, 1}-1} 
\sum_{ n' =0}^{N-1 }  w_{-1,1}(\nu') \sin
\frac{\pi ( 2\nu+1)(2\nu'+2)}{\bf k}
 \omega^{n(n'-\frac{1}{2})}  |\nu'>_{-1, 1 } 
\otimes |n'>  \ , \\
{\rm S}(|\nu>_{-1, 1} \otimes |n>)  &= 
\frac{4 }
{\bf k} ( \sum_{ \nu' =0}^{M_{1,-1}-1} 
\sum_{ n' =0}^{N-1 } 
\sin \frac{\pi (2\nu +2)(2\nu'+1)}{\bf k}
\omega^{(n-\frac{1}{2})n'}  
|\nu'>_{1,-1}  \otimes |n'>   \ , 
 \\
{\rm S}(|\nu>_{-1,-1} \otimes |n> )  
&=
\frac{4}
{\bf k}  \sum_{ \nu' =0}^{M_{-1,-1}-1}
\sum_{ n' =0}^{N-1}  
\sin \frac{\pi (2\nu+2)(2\nu'+2)}{\bf k}
\omega^{(n-\frac{1}{2})(n'-\frac{1}{2})} 
|\nu'>_{-1,-1} \otimes |n'> \ .
\end{array}
$$
Hence we obtain the matrices of ${\rm S}$
for $\Pi_{\rm NS}$. By Proposition 4 the 
${\rm S}$-expression of  a Ramon
sector differs from the one of the Neveu-Schwarz
sector only by a factor $e^{2 \pi
\sqrt{-1}(\pm Q_\lambda +\frac{\bf c}{4}) }$.
Therefore, with the equalities,
$$
\begin{array}{ll}
e^{2 \pi \sqrt{-1}(\pm Q_{[2\nu,-2n]} +\frac{\bf
c}{4}) } =\sqrt{-1}
\omega^{\mp n-1/4} \ , &  e^{2 \pi \sqrt{-1}(
Q_{[2\nu+1,1-2n]} +\frac{\bf c}{4}) } = \sqrt{-1}
\omega^{- n+1/4} \ , \\ 
e^{2 \pi \sqrt{-1}(-Q_{[2\nu+1,1-2n]} +\frac{\bf
c}{4}) }  =\sqrt{-1} \omega^{n-\frac{3}{4}} \ .
\end{array}
$$
the ${\rm S}$-transformation matrices for $\Pi_{\rm
R}, \Pi_{\widetilde{\rm R}}$  follow from that of
$\Pi_{\rm NS}$.
$\Box$ \par \vspace{.2in} \noindent 

\section{Acknowledgements}
I would like to thank MSRI and
Mathematical Department of U.C. Berkeley, U. S.
A. for the hospitality during the academic year
of 1997, where this work was formed. Part of this
work was performed during a visit of the author 
to the Hong Kong University of Science and
Technology (HKUST), Hong Kong on November, 1998. I
thank Professor Xiaoping Xu for his kind
invitation and hospitality at HKUST. 
 
\section{Appendix: Geometrical Realization of
N=2 Conformal Algebra}
The geometrical representation of
generators of Virasoro algebra in terms of vector
fields on $\CZ^*$ has been well-known in
literature, so is the N$=$1 conformal
algebra's generators as derivations of $\CZ^*$
with one Grassmann variable $\theta$,  (see for
instance 
\cite{CK}).
The super-vector field expressions of
N-extended conformal algebra for a general 
N have been treated in a similar way, and 
its  various forms could be found in
literature
\cite{GR}. In this
section, we are going to present a detailed 
geometrical realization of the N$= 2$
conformal algebra so that one could have a 
better understanding of the symmetries
represented by this algebra. We start with  an
alternative definition of the N$=$2 conformal
algebra, whose formulation works also for an
arbitrary  value N,   
$$
\begin{array}{ll}
[ L_m , L_n ] = (m-n) L_{m+n} + \frac{c}{12}(m^3-m) 
\delta_{m+n,0} \ , &
[ J_m , J_n ] = \frac{c}{3} m \delta_{m+n, 0} \ , 
\\
  
[L_m , J_n ] = - n J_{m+n} \ , \ \
[ L_m, G_p^j ] = ( \frac{m}{2}- p ) G_{m+p}^j \ ,
\ \ &  [ J_m, G_p^j ] = \sqrt{-1}
\epsilon_{j,k} G_{m+p}^k \ , \ \ \ 
 \\
\{ G_p^j, G_q^k \} = 2 \delta_{j,k}L_{p+q} +
\sqrt{-1} \epsilon_{j,k}  ( p-q)J_{p+q} + 
\frac{c}{3}(p^2 - \frac{1}{4}) \delta_{j,k}
\delta_{p+q, 0}  \ ,
\end{array}
$$
where $j,k =1,2$, and $\epsilon_{j,k}$ are 
antisymmetric with $\epsilon_{1,2}=1$. The
definition of ${\sf SCA}$ in Sect. 3 is equivalent
to the above one by the  relations,
$$ 
G^+_{n}= \frac{1}{\sqrt2} (G_n^1 +
\sqrt{-1}G_n^2) \ ,  \ \ G^-_{n}=
\frac{1}{\sqrt2} (G_n^1 - \sqrt{-1}G_n^2) \ . 
$$
It is easy to see that 
the correspondence,
\bea(l)
L_n \mapsto (-1)^{n+1} L_{-n} \ , \ \ 
J_n \mapsto (-1)^n J_{-n} \ , \ \ \
G^*_p \mapsto (-1)^{p-\frac{1}{2}} G^*_{-p} \ ,
\ \  ( * = 1, 2 , {\rm or} \  \ \pm ) \ , 
\elea(trans)
together with $c \mapsto -c$, give rise to 
an automorphism of the superalgebra ${\sf SCA}$. 
There is a 
finite dimensional Lie-superalgebra ${\sf g}$ 
invariant under the above automorphism, 
the even part ${\sf g^e}$ and  odd part ${\sf
g^o}$ are given by 
$$
\begin{array}{ll}
{\sf g^e} = < L_{-1} , L_0 , L_1 , J_0 >_{\CZ}  
&\simeq sl_2 \oplus \CZ J_0  \ , 
\\  
{\sf g^o} = <
G^1_{\frac{\pm 1}{2}} , G^2_{\frac{\pm 1}{2}}
>_{\CZ} & = < G^\pm_{\frac{1}{2}} ,
G^\pm_{\frac{-1}{2}} >_{\CZ}  \ . 

\end{array}
$$
Here the standard basis elements $  e , f, h $
of
$sl_2$ with $
[h , e ] = 2 e  , [h, f ] = -2 f  ,  [e , f ]
= h $, are expressed by 
$$
 e = L_1 \ , \ h = -2 L_0 \ , \ f = -L_{-1} \ .
$$
The representation of the even algebra ${\sf g^e}$
on the odd space ${\sf g^o}$  decomposes 
into two
copies of the canonical  2-dimensional
representation of
$sl_2$:
$$
[ h , G^*_{\frac{\pm1}{2}} ] = \pm G^*_{\frac{1}{2}}  \ , \ \   
[e , G^*_{\frac{1}{2}} ] = 0 \ , 
\ [ e , G^*_{\frac{-1}{2}} ] = G^*_{\frac{1}{2}} \ , \ \ 
[ f , G^*_{\frac{1}{2}} ] = G^*_{\frac{-1}{2}}  \ , \ \ 
[ f , G^*_{\frac{-1}{2}} ] = 0 \ , 
$$ 
for $* = 1, 2, $ or $\pm$; while the action of $J_0$ is given by
$$
[ J_0 , G^j_p] = \sqrt{-1} \epsilon_{j,k} G_p^k \
,
\
\ {\rm or } \ \ \ [J_0 , G^{\pm}_p ] = \pm
G^{\pm}_p \ .
$$
The product of odd generators are given by the following relation:
$$
\{ G^j_{\frac{1}{2}} , G^k_{\frac{1}{2}}  \} = 2
\delta_{j,k} e \ 
\ , 
\{ G^j_{\frac{-1}{2}} , G^k_{\frac{-1}{2}}  \} = - 2
\delta_{j,k} f \  , \ \ 
\{ G^j_{\frac{-1}{2}} , G^k_{\frac{1}{2}}  \} = -
\delta_{j,k} h - \sqrt{-1} \epsilon_{j,k} J_0 \ , 
$$
equivalently
$$
\begin{array}{l}
\{ G_p^+ , G_q^+ \} = \{ G_p^- , G_q^- \} = 0 ,  \\
\{ G^+_{\frac{1}{2}} ,  G^-_{\frac{1}{2}} \} =  2 e \ , \ \
\{  G^+_{\frac{-1}{2}} , G^-_{\frac{-1}{2}} \} =  - 2 f , \ \
\{  G^+_{\frac{-1}{2}} ,  G^-_{\frac{1}{2}} \} = -h  - J_0 \ , \ \ 
\{  G^-_{\frac{-1}{2}} , G^+_{\frac{1}{2}}  \} = -h  + J_0 \ .
\end{array}
$$
In conformal field theory, the operators $f, \frac{h}{2}, e, 
G^*_{\frac{-1}{2}}, G^*_{\frac{1}{2}}, J_0$
are called the momentum, dilation, special-conformal, supersymmetry, 
s-supersymmetry, and $SO(2)$-charge generators.
Let $\CZ^{* 1,2}$ be the one dimensional
superspace with the coordinate, $t \in \CZ^*$,
and  the two-component Grassmann variable,
$\theta=(\theta^1, \theta^2)$. The ${\sf SCA}$ 
with $c=0$ can be represented by the following 
super-vector fields 
on $\CZ^{* 1,2}$,
\bea(l)
L_n = - t^n( t \frac{\partial}{\partial t} + 
\frac{n+1}{2}  \theta^k \frac{
\partial}{\partial \theta^k} ) \ , 
\\ G^j_p = \sqrt{2\sqrt{-1}}
t^{p+\frac{1}{2}} [ 
\theta^j ( \frac{\partial}{\partial t} + 
\frac{2p+1}{2t}  
\theta^k \frac{
\partial}{\partial \theta^k})
- \frac{1}{2\sqrt{-1}}\frac{
\partial}{\partial \theta^j} ] \ , \\
J_n = \sqrt{-1} t^n ( \theta^1 
\frac{
\partial}{\partial \theta^2} -
\theta^2 \frac{
\partial}{\partial \theta^1} ) \ .
\elea(vrep)
Under the conformal transformation, 
$$
(t, \theta) \mapsto ( -t^{-1} , -t^{-1}
\theta ) \  , \ \ \theta= (\theta^1, \theta^2) \ ,
$$
one has the relations,
$$
\frac{\partial}{\partial t} \mapsto t^2
\frac{\partial}{\partial t} + t \theta^k
\frac{
\partial}{\partial \theta^k} \ , \ \ 
\frac{
\partial}{\partial \theta^j}
\mapsto - t \frac{
\partial}{\partial \theta^j} \ ,
$$
which give rise to the automorphism 
(\req(trans)) of ${\sf SCA}$. Let 
$\PZ^{1,2}$ be the one-dimensional 
projective superspace, which is 
the union of two affine 
spaces $U_0 , U_{\infty}$ with the
super-coordinates:
$$
(t, \theta ) \in \CZ^{1,2} \simeq U_0  \ , \ \ 
(\tilde{t}, \tilde{\theta}) 
\in
\CZ^{1,2}
\simeq U_{\infty } \ .
$$
The intersection, $U_{0} \cap U_{\infty}$,
consists of elements with 
$t \neq 0,$ or $\tilde{t} \neq 0$, with the
coordinate transformation
$$
(\tilde{t}, \tilde{\theta})  = ( -t^{-1} , -t^{-1}
\theta ) \in \CZ^{* 1,2} \ .
$$
We shall denote ${\rm Vect}(M)$ the space of
super-vector fields of a  super-manifold $M$. 
Then through (\req(vrep)), one can regard
$L_n, G^j_p, J_n$  as elements in  ${\rm
Vect}(U_0 \cap U_{\infty})$, and we have 
$$
\begin{array}{llll}
L_n \ \ \ (n \geq -1) , & G^*_p \ \ \ ( p \geq \frac{-1}{2}) , & 
J_m \ \ \ (m \geq 0) & \in {\rm Vect}(U_0) \ , \\
L_n \ \ \ (n \leq 1) , & G^*_p \ \ \ (p \leq \frac{1}{2}) , 
& J_m \ \ \ (m \leq 0) & \in {\rm Vect}(U_{\infty})
\end{array}
$$ 
Hence 
$$
{\sf g} = <L_n, J_m, G^j_p >_{\CZ} \cap {\rm
Vect}(\PZ^{1,2}) \ .
$$
Now one can clearly see the analogy of above
description of N=2 conformal algebra with 
the usual Virasoro interpretation of vector 
fields on 1-dimensional projective space $\PZ^1$.


\begin{thebibliography}{99}
\bibitem{BFK} W. Boucher, D. Friedan and A. Kent, Determinant formulae 
and unitarity for the  $N=2$ superconformal algebras in two dimensions 
or exact results on string compactification, 
Phys. Lett. B 172 (1987) 316-327.
\bibitem{CK} S.J. Cheng and  V. G. Kac, Conformal 
modules, Asian J. Math. 1 (1997) 181-193. 
\bibitem{D} V. K. Dobrev, Characters of the
unitarizable highest weight  modules over $N=2$
superconformal algebras, Phys. Lett. B 186
(1987)  43-51.
\bibitem{DG} M. D\"{o}zzrapf and B. Gato-Rivera,
Determinant formula for the topological N=2
superconformal algebra, Nucl. Phys. B558 (1999)
503, hep-th/9905063.
\bibitem{DVV} R. Dijkgraaf, E. Verlinde and H.
Verlinde, Topological strings in $d<1$, Nucl.
Phys. B352 (1991) 59.
\bibitem{EOTY} T. Eguchi, H. Ooguri, A. Taormina,
and S. K. Yang, Superconformal algebras and
string compactification on manifolds with SU(n)
holonomy, Nucl. Phys. B315 (1989) 193-221.
\bibitem{EZ} M. Eichler and D. Zagier, The
theory of Jacobi forms, Progress in Math.,
Birkh\"{a}user Boston, 1985.
\bibitem{FSST} B.L. Feigin, A.M. Semikhatov, V.A.
Sirota, and I. Yu. Tipunin, Resolutions and
characters of irreducible representations of the 
$N=2$ superconformal algebra, \ hep-th/9805179.
\bibitem{Gep} D. Gepner, Exactly solvable string
compactification on manifolds of SU(N) holonomy,
Phys. Lett. 199B (1987) 380; Space-time
supersymmetry in compactified string theory and
superconformal models, Nucl. Phys. B296 (1988) 757.
\bibitem{GR} S. J. Gates, Jr. and L. Rana, Superspace Geometrical 
representations of extended super Virasoro 
algebra, Phys. Lett. B 438 (1998)
80, hep-th/9806038.
\bibitem{HBJ} F. Hirzebruch, T. Berger and R. Jung, Manifolds and modular forms, 
Aspects of Mathematics, Friedr. Vieweg \&
Sohn Verlagsgesellschaft mbH,
Braunschweig/Wiesbaden, 1992. 

\bibitem{H} F. Hirzebruch, Complex cobordism and
the elliptic genus, (notes by A. Langer),
Contemporary Mathematics Vol. 241, 1999 American
Mathematical Society, 9-20.
\bibitem{Ka} V. G. Kac, Vertex algebra for
beginners, Providence: AMS,  University Lecture
Notes, vol. 9 (1996).
\bibitem{KYY} T. Kawai, Y. Yamada and S. K. Yang,
Elliptic genera and 
$N=2$ superconformal field theory, Nucl. Phys. B
414 (1994) 191-212, \  hep-th/9306096.
\bibitem{Ki} E. B. Kiritsis, Character formulae
and the structure of the presentations of the
$N=1, N=2$ superconformal algebras, Int. J. Mod.
Phys. A , 3 (1988) 1871-1906. 
\bibitem{M} Y. Matsuo, Character formula of $C<1$
unitary representation of $N=2$ superconformal
algebra, Prog. Theor. Phys. 77 (1987) 793-797.
\bibitem{RY} F. Ravanini and S-K Yang, Modular invariance in $N=2$ 
superconformal field theories, Phys. Lett. B195 (1987), 202-208.
\bibitem{R} S. S. Roan, Modular invariance of
manifolds with $SU(n)$  holonomy, Intern. J.
Math. Vol. 3, No. 3 (1992) 359-395.
\bibitem{R00} S. S. Roan, Elliptic genus and N=2
conformal algebra ( in preparation).
\bibitem{Wa} M. Wakimoto, Fusion rules for N=2
superconformal modules, hep-th/9807144. 
\bibitem{W} E. Witten, On the Landau-Ginzburg
description of $N=2$  minimal models, Int. J.
Mod. Phys. A9 (1994) 4783, \ hep-th/9304026.


\end{thebibliography}
\end{document}